\documentclass{eptcs}
\providecommand{\event}{TFPIE 2025}

\usepackage{subcaption}
\usepackage{amsmath}
\usepackage{amssymb}
\usepackage{hyperref}
\usepackage{cleveref}
\usepackage{alltt}
\usepackage{tikz} 
\usetikzlibrary{automata, positioning, arrows} 
\usepackage{float}
\floatstyle{ruled}
\restylefloat{figure}
\usepackage{mathtools}
\usepackage{dsfont}
\usepackage{pifont}
\usepackage{cancel}
\usepackage{enumitem}
\usepackage{amsfonts}
\usepackage{subcaption}
\usepackage{hyperref}
\usepackage{cleveref}
\usepackage{alltt}
\usepackage{tikz}
\usepackage{float}
\floatstyle{ruled}
\restylefloat{figure}
\usepackage{pgfplots}
\pgfplotsset{compat=1.18}
\usepackage{balance}
\usepackage{listings}

\usepackage[T1]{fontenc}

	\lstdefinelanguage{racket} {
		morecomment=[l]{;},         
		morecomment=[s]{\#|}{|\#},  
		commentstyle={\color{darkorange}\slshape\sffamily},
		morestring=[b]",
		stringstyle=\color{red},
		literate=%
		{->}{{$\rightarrow$}}1
		{*}{{*}}1
		{lambda}{{\textcolor{blue}{\(\lambda\)}}}1
		{EMP}{{\ep{}}}1,
		classoffset=1,
		morekeywords={
			define, define-syntax, define-macro, define-datatype, define-stream, \#lang, stream-lambda
		},
		keywordstyle=\color{purple},
		classoffset=2,
		morekeywords={
			*, *list/c, +, -, /, <, </c, <=, =, >, >/c, >=, abort-current-continuation, abs, absolute-path?, acos, add1, alarm-evt, always-evt, andmap, angle, append, append*, append-map, argmax, argmin, arithmetic-shift, arity-at-least-value, arity-at-least?, arity-checking-wrapper, arity-includes?, arity=?, arrow-contract-info-accepts-arglist, arrow-contract-info-chaperone-procedure, arrow-contract-info-check-first-order, arrow-contract-info?, asin, assert-unreachable, assf, assoc, assq, assv, assw, atan, banner, base->-doms/c, base->-rngs/c, base->?, bitwise-and, bitwise-bit-field, bitwise-bit-set?, bitwise-ior, bitwise-not, bitwise-xor, blame-add-car-context, blame-add-cdr-context, blame-add-missing-party, blame-add-nth-arg-context, blame-add-range-context, blame-add-unknown-context, blame-context, blame-contract, blame-fmt->-string, blame-missing-party?, blame-negative, blame-original?, blame-positive, blame-replace-negative, blame-replaced-negative?, blame-source, blame-swap, blame-swapped?, blame-update, blame-value, blame?, block-device-type-bits, boolean=?, boolean?, bound-identifier=?, box, box-cas!, box-immutable, box?, break-enabled, break-parameterization?, break-thread, build-chaperone-contract-property, build-compound-type-name, build-contract-property, build-flat-contract-property, build-list, build-path, build-path/convention-type, build-string, build-vector, byte-pregexp, byte-pregexp?, byte-ready?, byte-regexp, byte-regexp?, byte?, bytes, bytes->immutable-bytes, bytes->list, bytes->path, bytes->path-element, bytes->string/latin-1, bytes->string/locale, bytes->string/utf-8, bytes-append, bytes-append*, bytes-close-converter, bytes-convert, bytes-convert-end, bytes-converter?, bytes-copy, bytes-copy!, bytes-environment-variable-name?, bytes-fill!, bytes-join, bytes-length, bytes-no-nuls?, bytes-open-converter, bytes-ref, bytes-set!, bytes-utf-8-index, bytes-utf-8-length, bytes-utf-8-ref, bytes<?, bytes=?, bytes>?, bytes?, caaaar, caaadr, caaar, caadar, caaddr, caadr, caar, cadaar, cadadr, cadar, caddar, cadddr, caddr, cadr, call-in-continuation, call-in-nested-thread, call-with-break-parameterization, call-with-composable-continuation, call-with-continuation-barrier, call-with-continuation-prompt, call-with-current-continuation, call-with-default-reading-parameterization, call-with-escape-continuation, call-with-exception-handler, call-with-immediate-continuation-mark, call-with-input-bytes, call-with-input-string, call-with-output-bytes, call-with-output-string, call-with-parameterization, call-with-semaphore, call-with-semaphore/enable-break, call-with-values, call/cc, call/ec, car, cartesian-product, cdaaar, cdaadr, cdaar, cdadar, cdaddr, cdadr, cdar, cddaar, cddadr, cddar, cdddar, cddddr, cdddr, cddr, cdr, ceiling, channel-get, channel-put, channel-put-evt, channel-put-evt?, channel-try-get, channel?, chaperone-box, chaperone-channel, chaperone-continuation-mark-key, chaperone-contract-property?, chaperone-contract?, chaperone-evt, chaperone-hash, chaperone-hash-set, chaperone-of?, chaperone-procedure, chaperone-procedure*, chaperone-prompt-tag, chaperone-struct, chaperone-struct-type, chaperone-vector, chaperone-vector*, chaperone?, char->integer, char-alphabetic?, char-blank?, char-ci<=?, char-ci<?, char-ci=?, char-ci>=?, char-ci>?, char-downcase, char-extended-pictographic?, char-foldcase, char-general-category, char-grapheme-break-property, char-grapheme-step, char-graphic?, char-in, char-iso-control?, char-lower-case?, char-numeric?, char-punctuation?, char-ready?, char-symbolic?, char-title-case?, char-titlecase, char-upcase, char-upper-case?, char-utf-8-length, char-whitespace?, char<=?, char<?, char=?, char>=?, char>?, char?, character-device-type-bits, check-duplicate-identifier, checked-procedure-check-and-extract, choice-evt, class->interface, class-info, class-seal, class-unseal, class?, cleanse-path, close-input-port, close-output-port, coerce-chaperone-contract, coerce-chaperone-contracts, coerce-contract, coerce-contract/f, coerce-contracts, coerce-flat-contract, coerce-flat-contracts, collect-garbage, collection-file-path, collection-path, combinations, combine-output, compile, compile-allow-set!-undefined, compile-context-preservation-enabled, compile-enforce-module-constants, compile-syntax, compile-target-machine?, compiled-expression-add-target-machine, compiled-expression-recompile, compiled-expression?, compiled-module-expression?, complete-path?, complex/c, complex?, compose, compose1, conjoin, conjugate, cons, cons?, const, const*, continuation-mark-key?, continuation-mark-set->context, continuation-mark-set->iterator, continuation-mark-set->list, continuation-mark-set->list*, continuation-mark-set-first, continuation-mark-set?, continuation-marks, continuation-prompt-available?, continuation-prompt-tag?, continuation?, contract-continuation-mark-key, contract-custom-write-property-proc, contract-equivalent?, contract-first-order, contract-first-order-passes?, contract-late-neg-projection, contract-name, contract-proc, contract-projection, contract-property?, contract-random-generate, contract-random-generate-env?, contract-random-generate-fail, contract-random-generate-fail?, contract-random-generate-get-current-environment, contract-random-generate-stash, contract-random-generate/choose, contract-stronger?, contract-struct-exercise, contract-struct-generate, contract-struct-late-neg-projection, contract-struct-list-contract?, contract-val-first-projection, contract?, convert-stream, copy-file, copy-port, cos, cosh, count, current-blame-format, current-break-parameterization, current-code-inspector, current-command-line-arguments, current-compile, current-compile-realm, current-compile-target-machine, current-compiled-file-roots, current-continuation-marks, current-custodian, current-directory, current-directory-for-user, current-drive, current-environment-variables, current-error-message-adjuster, current-error-port, current-eval, current-evt-pseudo-random-generator, current-force-delete-permissions, current-future, current-gc-milliseconds, current-get-interaction-evt, current-get-interaction-input-port, current-inexact-milliseconds, current-inexact-monotonic-milliseconds, current-input-port, current-inspector, current-library-collection-links, current-library-collection-paths, current-load, current-load-extension, current-load-relative-directory, current-load/use-compiled, current-locale, current-logger, current-memory-use, current-milliseconds, current-module-declare-name, current-module-declare-source, current-module-name-resolver, current-module-path-for-load, current-namespace, current-output-port, current-parameterization, current-plumber, current-preserved-thread-cell-values, current-print, current-process-milliseconds, current-prompt-read, current-pseudo-random-generator, current-read-interaction, current-reader-guard, current-readtable, current-seconds, current-security-guard, current-subprocess-custodian-mode, current-subprocess-keep-file-descriptors, current-thread, current-thread-group, current-thread-initial-stack-size, current-write-relative-directory, curry, curryr, custodian-box-value, custodian-box?, custodian-limit-memory, custodian-managed-list, custodian-memory-accounting-available?, custodian-require-memory, custodian-shut-down?, custodian-shutdown-all, custodian?, custom-print-quotable-accessor, custom-print-quotable?, custom-write-accessor, custom-write-property-proc, custom-write?, date*-nanosecond, date*-time-zone-name, date*?, date-day, date-dst?, date-hour, date-minute, date-month, date-second, date-time-zone-offset, date-week-day, date-year, date-year-day, date?, datum->syntax, datum-intern-literal, default-continuation-prompt-tag, default-global-port-print-handler, degrees->radians, delete-directory, delete-file, denominator, dict-iter-contract, dict-key-contract, dict-value-contract, directory-exists?, directory-list, directory-type-bits, disjoin, display, displayln, double-flonum?, drop, drop-common-prefix, drop-right, dropf, dropf-right, dump-memory-stats, dup-input-port, dup-output-port, dynamic-get-field, dynamic-object/c, dynamic-require, dynamic-require-for-syntax, dynamic-send, dynamic-set-field!, dynamic-wind, eighth, empty, empty-sequence, empty-stream, empty?, environment-variables-copy, environment-variables-names, environment-variables-ref, environment-variables-set!, environment-variables?, eof, eof-object?, ephemeron-value, ephemeron?, eprintf, eq-contract-val, eq-contract?, eq-hash-code, eq?, equal-always-hash-code, equal-always-hash-code/recur, equal-always-secondary-hash-code, equal-always?, equal-always?/recur, equal-contract-val, equal-contract?, equal-hash-code, equal-hash-code/recur, equal-secondary-hash-code, equal<\%>, equal?, equal?/recur, eqv-hash-code, eqv?, error, error-contract->adjusted-string, error-display-handler, error-escape-handler, error-message->adjusted-string, error-message-adjuster-key, error-print-context-length, error-print-source-location, error-print-width, error-syntax->string-handler, error-value->string-handler, eval, eval-jit-enabled, eval-syntax, even?, evt/c, evt?, exact->inexact, exact-ceiling, exact-floor, exact-integer?, exact-nonnegative-integer?, exact-positive-integer?, exact-round, exact-truncate, exact?, executable-yield-handler, exit, exit-handler, exn-continuation-marks, exn-message, exn:break-continuation, exn:break:hang-up?, exn:break:terminate?, exn:break?, exn:fail:contract:arity?, exn:fail:contract:blame-object, exn:fail:contract:blame?, exn:fail:contract:continuation?, exn:fail:contract:divide-by-zero?, exn:fail:contract:non-fixnum-result?, exn:fail:contract:variable-id, exn:fail:contract:variable?, exn:fail:contract?, exn:fail:filesystem:errno-errno, exn:fail:filesystem:errno?, exn:fail:filesystem:exists?, exn:fail:filesystem:missing-module-path, exn:fail:filesystem:missing-module?, exn:fail:filesystem:version?, exn:fail:filesystem?, exn:fail:network:errno-errno, exn:fail:network:errno?, exn:fail:network?, exn:fail:object?, exn:fail:out-of-memory?, exn:fail:read-srclocs, exn:fail:read:eof?, exn:fail:read:non-char?, exn:fail:read?, exn:fail:syntax-exprs, exn:fail:syntax:missing-module-path, exn:fail:syntax:missing-module?, exn:fail:syntax:unbound?, exn:fail:syntax?, exn:fail:unsupported?, exn:fail:user?, exn:fail?, exn:misc:match?, exn:missing-module-accessor, exn:missing-module?, exn:srclocs-accessor, exn:srclocs?, exn?, exp, expand, expand-once, expand-syntax, expand-syntax-once, expand-syntax-to-top-form, expand-to-top-form, expand-user-path, explode-path, expt, externalizable<\%>, failure-result/c, false, false/c, false?, field-names, fifo-type-bits, fifth, file-exists?, file-name-from-path, file-or-directory-identity, file-or-directory-modify-seconds, file-or-directory-permissions, file-or-directory-stat, file-or-directory-type, file-position, file-position*, file-size, file-stream-buffer-mode, file-stream-port?, file-truncate, file-type-bits, filename-extension, filesystem-change-evt, filesystem-change-evt-cancel, filesystem-change-evt?, filesystem-root-list, filter, filter-map, filter-not, filter-read-input-port, find-compiled-file-roots, find-executable-path, find-library-collection-links, find-library-collection-paths, find-system-path, findf, first, fixnum?, flat-contract, flat-contract-predicate, flat-contract-property?, flat-contract?, flat-named-contract, flatten, floating-point-bytes->real, flonum?, floor, flush-output, fold-files, foldl, foldr, for-each, force, format, fourth, fprintf, free-identifier=?, free-label-identifier=?, free-template-identifier=?, free-transformer-identifier=?, fsemaphore-count, fsemaphore-post, fsemaphore-try-wait?, fsemaphore-wait, fsemaphore?, future, future?, futures-enabled?, gcd, generate-member-key, generate-temporaries, generic-set?, generic?, gensym, get-output-bytes, get-output-string, get/build-late-neg-projection, get/build-val-first-projection, getenv, global-port-print-handler, group-by, group-execute-bit, group-permission-bits, group-read-bit, group-write-bit, guard-evt, handle-evt, handle-evt?, has-blame?, has-contract?, hash, hash->list, hash-clear, hash-clear!, hash-copy, hash-count, hash-empty?, hash-ephemeron?, hash-eq?, hash-equal-always?, hash-equal?, hash-eqv?, hash-for-each, hash-has-key?, hash-iterate-first, hash-iterate-key, hash-iterate-key+value, hash-iterate-next, hash-iterate-pair, hash-iterate-value, hash-keys, hash-keys-subset?, hash-map, hash-placeholder?, hash-ref, hash-ref!, hash-ref-key, hash-remove, hash-remove!, hash-set, hash-set!, hash-set*, hash-set*!, hash-strong?, hash-update, hash-update!, hash-values, hash-weak?, hash?, hashalw, hasheq, hasheqv, identifier-binding, identifier-binding-portal-syntax, identifier-binding-symbol, identifier-distinct-binding, identifier-label-binding, identifier-prune-lexical-context, identifier-prune-to-source-module, identifier-remove-from-definition-context, identifier-template-binding, identifier-transformer-binding, identifier?, identity, if/c, imag-part, immutable?, impersonate-box, impersonate-channel, impersonate-continuation-mark-key, impersonate-hash, impersonate-hash-set, impersonate-procedure, impersonate-procedure*, impersonate-prompt-tag, impersonate-struct, impersonate-vector, impersonate-vector*, impersonator-contract?, impersonator-ephemeron, impersonator-of?, impersonator-prop:application-mark, impersonator-prop:blame, impersonator-prop:contracted, impersonator-property-accessor-procedure?, impersonator-property?, impersonator?, implementation?, implementation?/c, in-combinations, in-cycle, in-dict-pairs, in-parallel, in-permutations, in-sequences, in-values*-sequence, in-values-sequence, index-of, index-where, indexes-of, indexes-where, inexact->exact, inexact-real?, inexact?, infinite?, input-port?, inspector-superior?, inspector?, integer->char, integer->integer-bytes, integer-bytes->integer, integer-length, integer-sqrt, integer-sqrt/remainder, integer?, interface->method-names, interface-extension?, interface?, internal-definition-context-add-scopes, internal-definition-context-binding-identifiers, internal-definition-context-introduce, internal-definition-context-seal, internal-definition-context-splice-binding-identifier, internal-definition-context?, is-a?, is-a?/c, keyword->string, keyword-apply, keyword-apply/dict, keyword<?, keyword?, keywords-match, kill-thread, last, last-pair, lcm, length, liberal-define-context?, link-exists?, list, list*, list->bytes, list->mutable-set, list->mutable-setalw, list->mutable-seteq, list->mutable-seteqv, list->set, list->setalw, list->seteq, list->seteqv, list->string, list->vector, list->weak-set, list->weak-setalw, list->weak-seteq, list->weak-seteqv, list-contract?, list-prefix?, list-ref, list-set, list-tail, list-update, list?, listen-port-number?, load, load-extension, load-on-demand-enabled, load-relative, load-relative-extension, load/cd, load/use-compiled, local-expand, local-expand/capture-lifts, local-transformer-expand, local-transformer-expand/capture-lifts, locale-string-encoding, log, log-all-levels, log-level-evt, log-level?, log-max-level, log-message, log-receiver?, logger-name, logger?, magnitude, make-arity-at-least, make-base-empty-namespace, make-base-namespace, make-bytes, make-channel, make-chaperone-contract, make-continuation-mark-key, make-continuation-prompt-tag, make-contract, make-custodian, make-custodian-box, make-date, make-date*, make-derived-parameter, make-directory, make-directory*, make-do-sequence, make-empty-namespace, make-environment-variables, make-ephemeron, make-ephemeron-hash, make-ephemeron-hashalw, make-ephemeron-hasheq, make-ephemeron-hasheqv, make-exn, make-exn:break, make-exn:break:hang-up, make-exn:break:terminate, make-exn:fail, make-exn:fail:contract, make-exn:fail:contract:arity, make-exn:fail:contract:blame, make-exn:fail:contract:continuation, make-exn:fail:contract:divide-by-zero, make-exn:fail:contract:non-fixnum-result, make-exn:fail:contract:variable, make-exn:fail:filesystem, make-exn:fail:filesystem:errno, make-exn:fail:filesystem:exists, make-exn:fail:filesystem:missing-module, make-exn:fail:filesystem:version, make-exn:fail:network, make-exn:fail:network:errno, make-exn:fail:object, make-exn:fail:out-of-memory, make-exn:fail:read, make-exn:fail:read:eof, make-exn:fail:read:non-char, make-exn:fail:syntax, make-exn:fail:syntax:missing-module, make-exn:fail:syntax:unbound, make-exn:fail:unsupported, make-exn:fail:user, make-file-or-directory-link, make-flat-contract, make-fsemaphore, make-generic, make-hash, make-hash-placeholder, make-hashalw, make-hashalw-placeholder, make-hasheq, make-hasheq-placeholder, make-hasheqv, make-hasheqv-placeholder, make-immutable-hash, make-immutable-hashalw, make-immutable-hasheq, make-immutable-hasheqv, make-impersonator-property, make-input-port, make-input-port/read-to-peek, make-inspector, make-interned-syntax-introducer, make-keyword-procedure, make-known-char-range-list, make-list, make-lock-file-name, make-log-receiver, make-logger, make-mixin-contract, make-none/c, make-output-port, make-parameter, make-parent-directory*, make-phantom-bytes, make-pipe, make-pipe-with-specials, make-placeholder, make-plumber, make-polar, make-portal-syntax, make-prefab-struct, make-primitive-class, make-proj-contract, make-pseudo-random-generator, make-reader-graph, make-readtable, make-rectangular, make-rename-transformer, make-resolved-module-path, make-security-guard, make-semaphore, make-set!-transformer, make-shared-bytes, make-sibling-inspector, make-special-comment, make-srcloc, make-string, make-struct-field-accessor, make-struct-field-mutator, make-struct-type, make-struct-type-property, make-syntax-delta-introducer, make-syntax-introducer, make-tentative-pretty-print-output-port, make-thread-cell, make-thread-group, make-vector, make-weak-box, make-weak-hash, make-weak-hashalw, make-weak-hasheq, make-weak-hasheqv, make-will-executor, map, match-equality-test, matches-arity-exactly?, max, mcar, mcdr, mcons, member, member-name-key-hash-code, member-name-key=?, member-name-key?, memf, memory-order-acquire, memory-order-release, memq, memv, memw, merge-input, method-in-interface?, min, mixin-contract, module->exports, module->imports, module->indirect-exports, module->language-info, module->namespace, module->realm, module-cache-clear!, module-compiled-cross-phase-persistent?, module-compiled-exports, module-compiled-imports, module-compiled-indirect-exports, module-compiled-language-info, module-compiled-name, module-compiled-realm, module-compiled-submodules, module-declared?, module-path-index-join, module-path-index-resolve, module-path-index-split, module-path-index-submodule, module-path-index?, module-path?, module-predefined?, module-provide-protected?, modulo, mpair?, mutable-set, mutable-setalw, mutable-seteq, mutable-seteqv, n->th, nack-guard-evt, namespace-anchor->empty-namespace, namespace-anchor->namespace, namespace-anchor?, namespace-attach-module, namespace-attach-module-declaration, namespace-base-phase, namespace-call-with-registry-lock, namespace-mapped-symbols, namespace-module-identifier, namespace-module-registry, namespace-require, namespace-require/constant, namespace-require/copy, namespace-require/expansion-time, namespace-set-variable-value!, namespace-symbol->identifier, namespace-syntax-introduce, namespace-undefine-variable!, namespace-unprotect-module, namespace-variable-value, namespace?, nan?, natural-number/c, natural?, negate, negative-integer?, negative?, never-evt, newline, ninth, non-empty-string?, nonnegative-integer?, nonpositive-integer?, normal-case-path, normalize-arity, normalize-path, normalized-arity?, not, null, null?, number->string, number?, numerator, object\%, object->vector, object-info, object-interface, object-method-arity-includes?, object-name, object-or-false=?, object=-hash-code, object=?, object?, odd?, open-input-bytes, open-input-string, open-output-bytes, open-output-nowhere, open-output-string, order-of-magnitude, ormap, other-execute-bit, other-permission-bits, other-read-bit, other-write-bit, output-port?, pair?, parameter-procedure=?, parameter?, parameterization?, parse-command-line, partition, path->bytes, path->complete-path, path->directory-path, path->string, path-add-extension, path-add-suffix, path-convention-type, path-element->bytes, path-element->string, path-element?, path-for-some-system?, path-get-extension, path-has-extension?, path-list-string->path-list, path-only, path-replace-extension, path-replace-suffix, path-string?, path<?, path?, peek-byte, peek-byte-or-special, peek-bytes, peek-bytes!, peek-bytes-avail!, peek-bytes-avail!*, peek-bytes-avail!/enable-break, peek-char, peek-char-or-special, peek-string, peek-string!, permutations, phantom-bytes?, pi, pi.f, pipe-content-length, place-break, place-channel, place-channel-get, place-channel-put, place-channel-put/get, place-channel?, place-dead-evt, place-enabled?, place-kill, place-location?, place-message-allowed?, place-wait, place?, placeholder-get, placeholder-set!, placeholder?, plumber-add-flush!, plumber-flush-all, plumber-flush-handle-remove!, plumber-flush-handle?, plumber?, poll-guard-evt, port->list, port-closed-evt, port-closed?, port-commit-peeked, port-count-lines!, port-count-lines-enabled, port-counts-lines?, port-display-handler, port-file-identity, port-file-unlock, port-next-location, port-number?, port-print-handler, port-progress-evt, port-provides-progress-evts?, port-read-handler, port-try-file-lock?, port-waiting-peer?, port-write-handler, port-writes-atomic?, port-writes-special?, port?, portal-syntax-content, portal-syntax?, positive-integer?, positive?, predicate/c, prefab-key->struct-type, prefab-key?, prefab-struct-key, prefab-struct-type-key+field-count, preferences-lock-file-mode, pregexp, pregexp-quote, pregexp?, pretty-display, pretty-print, pretty-print-.-symbol-without-bars, pretty-print-abbreviate-read-macros, pretty-print-columns, pretty-print-current-style-table, pretty-print-depth, pretty-print-exact-as-decimal, pretty-print-extend-style-table, pretty-print-handler, pretty-print-newline, pretty-print-post-print-hook, pretty-print-pre-print-hook, pretty-print-print-hook, pretty-print-print-line, pretty-print-remap-stylable, pretty-print-show-inexactness, pretty-print-size-hook, pretty-print-style-table?, pretty-printing, pretty-write, primitive-closure?, primitive-result-arity, primitive?, print, print-as-expression, print-boolean-long-form, print-box, print-graph, print-hash-table, print-mpair-curly-braces, print-pair-curly-braces, print-reader-abbreviations, print-struct, print-syntax-width, print-unreadable, print-value-columns, print-vector-length, printable/c, printable<\%>, printf, println, procedure->method, procedure-arity, procedure-arity-includes?, procedure-arity-mask, procedure-arity?, procedure-closure-contents-eq?, procedure-extract-target, procedure-impersonator*?, procedure-keywords, procedure-realm, procedure-reduce-arity, procedure-reduce-arity-mask, procedure-reduce-keyword-arity, procedure-reduce-keyword-arity-mask, procedure-rename, procedure-result-arity, procedure-specialize, procedure-struct-type?, procedure?, processor-count, progress-evt?, promise-forced?, promise-running?, promise/name?, promise?, prop:arity-string, prop:arrow-contract, prop:arrow-contract-get-info, prop:arrow-contract?, prop:authentic, prop:blame, prop:chaperone-contract, prop:checked-procedure, prop:contract, prop:contracted, prop:custom-print-quotable, prop:custom-write, prop:dict, prop:equal+hash, prop:evt, prop:exn:missing-module, prop:exn:srclocs, prop:expansion-contexts, prop:flat-contract, prop:impersonator-of, prop:input-port, prop:liberal-define-context, prop:object-name, prop:orc-contract, prop:orc-contract-get-subcontracts, prop:orc-contract?, prop:output-port, prop:place-location, prop:procedure, prop:recursive-contract, prop:recursive-contract-unroll, prop:recursive-contract?, prop:rename-transformer, prop:sealed, prop:sequence, prop:set!-transformer, prop:stream, proper-subset?, pseudo-random-generator->vector, pseudo-random-generator-vector?, pseudo-random-generator?, put-preferences, putenv, quotient, quotient/remainder, radians->degrees, raise, raise-argument-error, raise-argument-error*, raise-arguments-error, raise-arguments-error*, raise-arity-error, raise-arity-error*, raise-arity-mask-error, raise-arity-mask-error*, raise-contract-error, raise-mismatch-error, raise-range-error, raise-range-error*, raise-result-arity-error, raise-result-arity-error*, raise-result-error, raise-result-error*, raise-type-error, raise-user-error, random, random-seed, rational?, rationalize, read, read-accept-bar-quote, read-accept-box, read-accept-compiled, read-accept-dot, read-accept-graph, read-accept-infix-dot, read-accept-lang, read-accept-quasiquote, read-accept-reader, read-byte, read-byte-or-special, read-bytes, read-bytes!, read-bytes-avail!, read-bytes-avail!*, read-bytes-avail!/enable-break, read-bytes-line, read-case-sensitive, read-cdot, read-char, read-char-or-special, read-curly-brace-as-paren, read-curly-brace-with-tag, read-decimal-as-inexact, read-eval-print-loop, read-installation-configuration-table, read-language, read-line, read-on-demand-source, read-single-flonum, read-square-bracket-as-paren, read-square-bracket-with-tag, read-string, read-string!, read-syntax, read-syntax-accept-graph, read-syntax/recursive, read/recursive, readtable-mapping, readtable?, real->decimal-string, real->double-flonum, real->floating-point-bytes, real->single-flonum, real-part, real?, reencode-input-port, reencode-output-port, regexp, regexp-match, regexp-match-exact?, regexp-match-peek, regexp-match-peek-immediate, regexp-match-peek-positions, regexp-match-peek-positions-immediate, regexp-match-peek-positions-immediate/end, regexp-match-peek-positions/end, regexp-match-positions, regexp-match-positions/end, regexp-match/end, regexp-match?, regexp-max-lookbehind, regexp-quote, regexp-replace, regexp-replace*, regexp-replace-quote, regexp-replaces, regexp-split, regexp-try-match, regexp?, regular-file-type-bits, relative-path?, remainder, remf, remf*, remove, remove*, remq, remq*, remv, remv*, remw, remw*, rename-contract, rename-file-or-directory, rename-transformer-target, rename-transformer?, replace-evt, reroot-path, resolve-path, resolved-module-path-name, resolved-module-path?, rest, reverse, round, second, seconds->date, security-guard?, semaphore-peek-evt, semaphore-peek-evt?, semaphore-post, semaphore-try-wait?, semaphore-wait, semaphore-wait/enable-break, semaphore?, sequence->list, sequence->stream, sequence-add-between, sequence-andmap, sequence-append, sequence-count, sequence-filter, sequence-fold, sequence-for-each, sequence-generate, sequence-generate*, sequence-length, sequence-map, sequence-ormap, sequence-ref, sequence-tail, sequence/c, sequence?, set, set!-transformer-procedure, set!-transformer?, set->list, set->stream, set-add, set-add!, set-box!, set-box*!, set-clear, set-clear!, set-copy, set-copy-clear, set-count, set-empty?, set-eq?, set-equal-always?, set-equal?, set-eqv?, set-first, set-for-each, set-group-id-bit, set-implements/c, set-implements?, set-intersect, set-intersect!, set-map, set-mcar!, set-mcdr!, set-member?, set-mutable?, set-phantom-bytes!, set-port-next-location!, set-remove, set-remove!, set-rest, set-subtract, set-subtract!, set-symmetric-difference, set-symmetric-difference!, set-union, set-union!, set-user-id-bit, set-weak?, set=?, set?, setalw, seteq, seteqv, seventh, sgn, sha1-bytes, sha224-bytes, sha256-bytes, shared-bytes, shell-execute, shrink-path-wrt, shuffle, simple-form-path, simplify-path, sin, single-flonum-available?, single-flonum?, sinh, sixth, skip-projection-wrapper?, sleep, socket-type-bits, some-system-path->string, special-comment-value, special-comment?, special-filter-input-port, split-at, split-at-right, split-common-prefix, split-path, splitf-at, splitf-at-right, sqr, sqrt, srcloc->string, srcloc-column, srcloc-line, srcloc-position, srcloc-source, srcloc-span, srcloc?, stencil-vector, stencil-vector-length, stencil-vector-mask, stencil-vector-mask-width, stencil-vector-ref, stencil-vector-set!, stencil-vector-update, stencil-vector?, sticky-bit, stop-after, stop-before, stream->list, stream-add-between, stream-andmap, stream-append, stream-count, stream-empty?, stream-filter, stream-first, stream-fold, stream-for-each, stream-force, stream-length, stream-map, stream-ormap, stream-ref, stream-rest, stream-tail, stream-take, stream/c, stream?, string, string->bytes/latin-1, string->bytes/locale, string->bytes/utf-8, string->immutable-string, string->keyword, string->list, string->number, string->path, string->path-element, string->some-system-path, string->symbol, string->uninterned-symbol, string->unreadable-symbol, string-append, string-append*, string-append-immutable, string-ci<=?, string-ci<?, string-ci=?, string-ci>=?, string-ci>?, string-contains?, string-copy, string-copy!, string-downcase, string-environment-variable-name?, string-fill!, string-foldcase, string-grapheme-count, string-grapheme-span, string-length, string-locale-ci<?, string-locale-ci=?, string-locale-ci>?, string-locale-downcase, string-locale-upcase, string-locale<?, string-locale=?, string-locale>?, string-no-nuls?, string-normalize-nfc, string-normalize-nfd, string-normalize-nfkc, string-normalize-nfkd, string-port?, string-prefix?, string-ref, string-set!, string-suffix?, string-titlecase, string-upcase, string-utf-8-length, string<=?, string<?, string=?, string>=?, string>?, string?, struct->vector, struct-accessor-procedure?, struct-constructor-procedure?, struct-info, struct-mutator-procedure?, struct-predicate-procedure?, struct-type-authentic?, struct-type-info, struct-type-make-constructor, struct-type-make-predicate, struct-type-property-accessor-procedure?, struct-type-property-predicate-procedure?, struct-type-property/c, struct-type-property?, struct-type-sealed?, struct-type?, struct:arity-at-least, struct:arrow-contract-info, struct:date, struct:date*, struct:exn, struct:exn:break, struct:exn:break:hang-up, struct:exn:break:terminate, struct:exn:fail, struct:exn:fail:contract, struct:exn:fail:contract:arity, struct:exn:fail:contract:blame, struct:exn:fail:contract:continuation, struct:exn:fail:contract:divide-by-zero, struct:exn:fail:contract:non-fixnum-result, struct:exn:fail:contract:variable, struct:exn:fail:filesystem, struct:exn:fail:filesystem:errno, struct:exn:fail:filesystem:exists, struct:exn:fail:filesystem:missing-module, struct:exn:fail:filesystem:version, struct:exn:fail:network, struct:exn:fail:network:errno, struct:exn:fail:object, struct:exn:fail:out-of-memory, struct:exn:fail:read, struct:exn:fail:read:eof, struct:exn:fail:read:non-char, struct:exn:fail:syntax, struct:exn:fail:syntax:missing-module, struct:exn:fail:syntax:unbound, struct:exn:fail:unsupported, struct:exn:fail:user, struct:srcloc, struct?, sub1, subbytes, subclass?, subclass?/c, subprocess, subprocess-group-enabled, subprocess-kill, subprocess-pid, subprocess-status, subprocess-wait, subprocess?, subset?, substring, suggest/c, symbol->string, symbol-interned?, symbol-unreadable?, symbol<?, symbol=?, symbol?, symbolic-link-type-bits, sync, sync/enable-break, sync/timeout, sync/timeout/enable-break, syntax->datum, syntax->list, syntax-arm, syntax-binding-set, syntax-binding-set->syntax, syntax-binding-set?, syntax-bound-phases, syntax-bound-symbols, syntax-column, syntax-debug-info, syntax-disarm, syntax-e, syntax-line, syntax-local-apply-transformer, syntax-local-bind-syntaxes, syntax-local-certifier, syntax-local-context, syntax-local-expand-expression, syntax-local-get-shadower, syntax-local-identifier-as-binding, syntax-local-introduce, syntax-local-lift-context, syntax-local-lift-expression, syntax-local-lift-module, syntax-local-lift-module-end-declaration, syntax-local-lift-provide, syntax-local-lift-require, syntax-local-lift-values-expression, syntax-local-make-definition-context, syntax-local-make-definition-context-introducer, syntax-local-make-delta-introducer, syntax-local-module-defined-identifiers, syntax-local-module-exports, syntax-local-module-interned-scope-symbols, syntax-local-module-required-identifiers, syntax-local-name, syntax-local-phase-level, syntax-local-submodules, syntax-local-transforming-module-provides?, syntax-local-value, syntax-local-value/immediate, syntax-original?, syntax-position, syntax-property, syntax-property-preserved?, syntax-property-remove, syntax-property-symbol-keys, syntax-protect, syntax-rearm, syntax-recertify, syntax-shift-phase-level, syntax-source, syntax-source-module, syntax-span, syntax-taint, syntax-tainted?, syntax-track-origin, syntax-transforming-module-expression?, syntax-transforming-with-lifts?, syntax-transforming?, syntax?, system-big-endian?, system-idle-evt, system-language+country, system-library-subpath, system-path-convention-type, system-type, tail-marks-match?, take, take-common-prefix, take-right, takef, takef-right, tan, tanh, tcp-abandon-port, tcp-accept, tcp-accept-evt, tcp-accept-ready?, tcp-accept/enable-break, tcp-addresses, tcp-close, tcp-connect, tcp-connect/enable-break, tcp-listen, tcp-listener?, tcp-port?, tentative-pretty-print-port-cancel, tentative-pretty-print-port-transfer, tenth, terminal-port?, the-unsupplied-arg, third, thread, thread-cell-ref, thread-cell-set!, thread-cell-values?, thread-cell?, thread-dead-evt, thread-dead?, thread-group?, thread-receive, thread-receive-evt, thread-resume, thread-resume-evt, thread-rewind-receive, thread-running?, thread-send, thread-suspend, thread-suspend-evt, thread-try-receive, thread-wait, thread/suspend-to-kill, thread?, time-apply, touch, true, truncate, udp-addresses, udp-bind!, udp-bound?, udp-close, udp-connect!, udp-connected?, udp-multicast-interface, udp-multicast-join-group!, udp-multicast-leave-group!, udp-multicast-loopback?, udp-multicast-set-interface!, udp-multicast-set-loopback!, udp-multicast-set-ttl!, udp-multicast-ttl, udp-open-socket, udp-receive!, udp-receive!*, udp-receive!-evt, udp-receive!/enable-break, udp-receive-ready-evt, udp-send, udp-send*, udp-send-evt, udp-send-ready-evt, udp-send-to, udp-send-to*, udp-send-to-evt, udp-send-to/enable-break, udp-send/enable-break, udp-set-receive-buffer-size!, udp-set-ttl!, udp-ttl, udp?, unbox, unbox*, uncaught-exception-handler, unit?, unquoted-printing-string, unquoted-printing-string-value, unquoted-printing-string?, unspecified-dom, unsupplied-arg?, use-collection-link-paths, use-compiled-file-check, use-compiled-file-paths, use-user-specific-search-paths, user-execute-bit, user-permission-bits, user-read-bit, user-write-bit, value-blame, value-contract, values, variable-reference->empty-namespace, variable-reference->module-base-phase, variable-reference->module-declaration-inspector, variable-reference->module-path-index, variable-reference->module-source, variable-reference->namespace, variable-reference->phase, variable-reference->resolved-module-path, variable-reference-constant?, variable-reference-from-unsafe?, variable-reference?, vector, vector*-append, vector*-copy, vector*-extend, vector*-length, vector*-ref, vector*-set!, vector*-set/copy, vector->immutable-vector, vector->list, vector->pseudo-random-generator, vector->pseudo-random-generator!, vector->values, vector-append, vector-argmax, vector-argmin, vector-cas!, vector-copy, vector-copy!, vector-count, vector-drop, vector-drop-right, vector-empty?, vector-extend, vector-fill!, vector-filter, vector-filter-not, vector-immutable, vector-length, vector-map, vector-map!, vector-member, vector-memq, vector-memv, vector-ref, vector-set!, vector-set*!, vector-set-performance-stats!, vector-set/copy, vector-split-at, vector-split-at-right, vector-take, vector-take-right, vector?, version, void, void?, weak-box-value, weak-box?, weak-set, weak-setalw, weak-seteq, weak-seteqv, will-execute, will-executor?, will-register, will-try-execute, with-input-from-bytes, with-input-from-string, with-output-to-bytes, with-output-to-string, would-be-future, wrap-evt, writable<\%>, write, write-byte, write-bytes, write-bytes-avail, write-bytes-avail*, write-bytes-avail-evt, write-bytes-avail/enable-break, write-char, write-special, write-special-avail*, write-special-evt, write-string, writeln, xor, zero?, check-tail-contract, for-clause-syntax-protect, legacy-match-expander?, match-...-nesting, match-expander?, prop:legacy-match-expander, prop:match-expander, syntax-local-match-introduce, syntax-pattern-variable?, \#\%app, \#\%datum, \#\%declare, \#\%expression, \#\%module-begin, \#\%plain-app, \#\%plain-lambda, \#\%plain-module-begin, \#\%printing-module-begin, \#\%provide, \#\%require, \#\%stratified-body, \#\%top, \#\%top-interaction, \#\%variable-reference, ->, ->*, ->*m, ->d, ->dm, ->i, ->m, ..., :do-in, <=/c, =/c, ==, =>, >=/c, _, absent, abstract, add-between, all-defined-out, all-from-out, and, and/c, any, any/c, apply, arity-at-least, arrow-contract-info, augment, augment*, augment-final, augment-final*, augride, augride*, bad-number-of-results, begin, begin-for-syntax, begin0, between/c, blame-add-context, box-immutable/c, box/c, call-with-atomic-output-file, call-with-file-lock/timeout, call-with-input-file, call-with-input-file*, call-with-output-file, call-with-output-file*, case, case->, case->m, case-lambda, channel/c, char-in/c, check-duplicates, class, class*, class-field-accessor, class-field-mutator, class/c, class/derived, combine-in, combine-out, command-line, compound-unit, compound-unit/infer, cond, cons/c, cons/dc, continuation-mark-key/c, contract, contract-exercise, contract-in, contract-out, contract-pos/neg-doubling, contract-struct, contracted, copy-directory/files, current-contract-region, date, date*, define, define-compound-unit, define-compound-unit/infer, define-contract-struct, define-custom-hash-types, define-custom-set-types, define-for-syntax, define-local-member-name, define-logger, define-match-expander, define-member-name, define-module-boundary-contract, define-namespace-anchor, define-opt/c, define-sequence-syntax, define-serializable-class, define-serializable-class*, define-signature, define-signature-form, define-splicing-for-clause-syntax, define-struct, define-struct/contract, define-struct/derived, define-syntax, define-syntax-rule, define-syntaxes, define-unit, define-unit-binding, define-unit-from-context, define-unit/contract, define-unit/new-import-export, define-unit/s, define-values, define-values-for-export, define-values-for-syntax, define-values/invoke-unit, define-values/invoke-unit/infer, define/augment, define/augment-final, define/augride, define/contract, define/final-prop, define/match, define/overment, define/override, define/override-final, define/private, define/public, define/public-final, define/pubment, define/subexpression-pos-prop, define/subexpression-pos-prop/name, delay, delay/idle, delay/name, delay/strict, delay/sync, delay/thread, delete-directory/files, dict->list, dict-can-functional-set?, dict-can-remove-keys?, dict-clear, dict-clear!, dict-copy, dict-count, dict-empty?, dict-for-each, dict-has-key?, dict-implements/c, dict-implements?, dict-iterate-first, dict-iterate-key, dict-iterate-next, dict-iterate-value, dict-keys, dict-map, dict-map/copy, dict-mutable?, dict-ref, dict-ref!, dict-remove, dict-remove!, dict-set, dict-set!, dict-set*, dict-set*!, dict-update, dict-update!, dict-values, dict?, display-lines, display-lines-to-file, display-to-file, do, dynamic->*, dynamic-instantiate, dynamic-place, dynamic-place*, else, eof-evt, except, except-in, except-out, exn, exn:break, exn:break:hang-up, exn:break:terminate, exn:fail, exn:fail:contract, exn:fail:contract:arity, exn:fail:contract:blame, exn:fail:contract:continuation, exn:fail:contract:divide-by-zero, exn:fail:contract:non-fixnum-result, exn:fail:contract:variable, exn:fail:filesystem, exn:fail:filesystem:errno, exn:fail:filesystem:exists, exn:fail:filesystem:missing-module, exn:fail:filesystem:version, exn:fail:network, exn:fail:network:errno, exn:fail:object, exn:fail:out-of-memory, exn:fail:read, exn:fail:read:eof, exn:fail:read:non-char, exn:fail:syntax, exn:fail:syntax:missing-module, exn:fail:syntax:unbound, exn:fail:unsupported, exn:fail:user, export, extends, failure-cont, field, field-bound?, file, file->bytes, file->bytes-lines, file->lines, file->list, file->string, file->value, find-files, find-relative-path, first-or/c, flat-contract-with-explanation, flat-murec-contract, flat-rec-contract, for, for*, for*/and, for*/async, for*/first, for*/fold, for*/fold/derived, for*/foldr, for*/foldr/derived, for*/hash, for*/hashalw, for*/hasheq, for*/hasheqv, for*/last, for*/list, for*/list/concurrent, for*/lists, for*/mutable-set, for*/mutable-setalw, for*/mutable-seteq, for*/mutable-seteqv, for*/or, for*/product, for*/set, for*/setalw, for*/seteq, for*/seteqv, for*/stream, for*/sum, for*/vector, for*/weak-set, for*/weak-setalw, for*/weak-seteq, for*/weak-seteqv, for-label, for-meta, for-space, for-syntax, for-template, for/and, for/async, for/first, for/fold, for/fold/derived, for/foldr, for/foldr/derived, for/hash, for/hashalw, for/hasheq, for/hasheqv, for/last, for/list, for/list/concurrent, for/lists, for/mutable-set, for/mutable-setalw, for/mutable-seteq, for/mutable-seteqv, for/or, for/product, for/set, for/setalw, for/seteq, for/seteqv, for/stream, for/sum, for/vector, for/weak-set, for/weak-setalw, for/weak-seteq, for/weak-seteqv, gen:custom-write, gen:dict, gen:equal+hash, gen:equal-mode+hash, gen:set, gen:stream, generic, get-field, get-preference, hash-copy-clear, hash-map/copy, hash/c, hash/dc, if, implies, import, in-bytes, in-bytes-lines, in-dict, in-dict-keys, in-dict-values, in-directory, in-ephemeron-hash, in-ephemeron-hash-keys, in-ephemeron-hash-pairs, in-ephemeron-hash-values, in-hash, in-hash-keys, in-hash-pairs, in-hash-values, in-immutable-hash, in-immutable-hash-keys, in-immutable-hash-pairs, in-immutable-hash-values, in-immutable-set, in-inclusive-range, in-indexed, in-input-port-bytes, in-input-port-chars, in-lines, in-list, in-mlist, in-mutable-hash, in-mutable-hash-keys, in-mutable-hash-pairs, in-mutable-hash-values, in-mutable-set, in-naturals, in-port, in-producer, in-range, in-set, in-slice, in-stream, in-string, in-syntax, in-value, in-vector, in-weak-hash, in-weak-hash-keys, in-weak-hash-pairs, in-weak-hash-values, in-weak-set, include, include-at/relative-to, include-at/relative-to/reader, include/reader, inclusive-range, inherit, inherit-field, inherit/inner, inherit/super, init, init-depend, init-field, init-rest, initiate-sequence, inner, input-port-append, inspect, instanceof/c, instantiate, integer-in, interface, interface*, invariant-assertion, invoke-unit, invoke-unit/infer, lambda, lazy, let, let*, let*-values, let-syntax, let-syntaxes, let-values, let/cc, let/ec, letrec, letrec-syntax, letrec-syntaxes, letrec-syntaxes+values, letrec-values, lib, link, list*of, list/c, listof, local, local-require, log-debug, log-error, log-fatal, log-info, log-warning, make-custom-hash, make-custom-hash-types, make-custom-set, make-custom-set-types, make-handle-get-preference-locked, make-immutable-custom-hash, make-limited-input-port, make-mutable-custom-set, make-object, make-temporary-directory, make-temporary-directory*, make-temporary-file, make-temporary-file*, make-weak-custom-hash, make-weak-custom-set, match, match*, match*/derived, match-define, match-define-values, match-lambda, match-lambda*, match-lambda**, match-let, match-let*, match-let*-values, match-let-values, match-letrec, match-letrec-values, match/derived, match/values, member-name-key, mixin, module, module*, module+, mutable-treelist/c, nand, new, new-∀/c, new-∃/c, non-empty-listof, none/c, nor, not/c, object-contract, object/c, one-of/c, only, only-in, only-meta-in, only-space-in, open, open-input-file, open-input-output-file, open-output-file, opt/c, or, or/c, overment, overment*, override, override*, override-final, override-final*, parameter/c, parameterize, parameterize*, parameterize-break, parametric->/c, pathlist-closure, peek-bytes!-evt, peek-bytes-avail!-evt, peek-bytes-evt, peek-string!-evt, peek-string-evt, peeking-input-port, place, place*, place/context, planet, port->bytes, port->bytes-lines, port->lines, port->string, prefix, prefix-in, prefix-out, pretty-format, private, private*, procedure-arity-includes/c, process, process*, process*/ports, process/ports, promise/c, prompt-tag/c, prop:dict/contract, property/c, protect-out, provide, provide-signature-elements, provide/contract, public, public*, public-final, public-final*, pubment, pubment*, quasiquote, quasisyntax, quasisyntax/loc, quote, quote-syntax, quote-syntax/prune, raise-blame-error, raise-not-cons-blame-error, raise-syntax-error, range, read-bytes!-evt, read-bytes-avail!-evt, read-bytes-evt, read-bytes-line-evt, read-line-evt, read-string!-evt, read-string-evt, real-in, recontract-out, recursive-contract, regexp-match*, regexp-match-evt, regexp-match-peek-positions*, regexp-match-positions*, relative-in, relocate-input-port, relocate-output-port, remove-duplicates, rename, rename-in, rename-inner, rename-out, rename-super, require, send, send*, send+, send-generic, send/apply, send/keyword-apply, set!, set!-values, set-field!, set/c, shared, sort, srcloc, stream, stream*, stream-cons, stream-lazy, string-join, string-len/c, string-normalize-spaces, string-replace, string-split, string-trim, struct, struct*, struct-copy, struct-field-index, struct-guard/c, struct-out, struct/c, struct/contract, struct/ctc, struct/dc, struct/derived, submod, super, super-instantiate, super-make-object, super-new, symbols, syntax, syntax-binding-set-extend, syntax-case, syntax-case*, syntax-deserialize, syntax-id-rules, syntax-rules, syntax-serialize, syntax/c, syntax/loc, system, system*, system*/exit-code, system/exit-code, tag, this, this\%, thunk, thunk*, time, transplant-input-port, transplant-output-port, treelist/c, unconstrained-domain->, unit, unit-from-context, unit/c, unit/new-import-export, unit/s, unless, unquote, unquote-splicing, unsyntax, unsyntax-splicing, values/drop, vector-immutable/c, vector-immutableof, vector-sort, vector-sort!, vector/c, vectorof, when, with-continuation-mark, with-contract, with-contract-continuation-mark, with-handlers, with-handlers*, with-input-from-file, with-method, with-output-to-file, with-syntax, write-to-file, ~.a, ~.s, ~.v, ~?, ~@, ~a, ~e, ~r, ~s, ~v, λ, equal?, equal, eq
		},
		keywordstyle=\color{blue},
		classoffset=3,
		morekeywords={import, export},
		keywordstyle=\color{green},
		classoffset=4,
		morekeywords={fsm, make-dfa, sm-apply, sm-states, sm-sigma, sm-start, sm-finals, sm-rules, sm-gamma, sm-showtransitions, sm-graph, gen-state, make-ndpda, make-ndfa, gen-regexp-word, singleton-regexp, kleenestar-regexp, concat-regexp, union-regexp, make-cfg, make-rg, make-csg, grammar-derive, grammar-not-derive, check-accept?, check-reject?, make-mttm},
		keywordstyle=\color{pakistangreen},
		classoffset=5,
		morekeywords={check-equal?, check-expect, check-equal, check-pred},
		keywordstyle=\color{sblue},
		classoffset=0,
		alsoletter={',`,-,/,>,<,\#,\%,?,=,*},
		moredelim=**[is][\color{lightgray}]{<<@<<}{>>@>>},
		moredelim=**[is][\itshape\color{OliveGreen}]{<<;<<}{>>;>>},
	}

\definecolor{sblue}{rgb}{0.14, 0.16, 0.48}
\definecolor{regalia}{rgb}{0.32, 0.18, 0.5}
\definecolor{palatinatepurple}{rgb}{0.41, 0.16, 0.38}
\definecolor{pakistangreen}{rgb}{0.0, 0.4, 0.0}
\definecolor{darkorange}{rgb}{1.0, 0.55, 0.0}

\newcommand{\racket}{\texttt{Racket}}

\newcommand{\rackunit}{\texttt{RackUnit}}
\newcommand{\fsm}{\texttt{FSM}}
\newcommand{\dfa}{\texttt{dfa}}
\newcommand{\ndfa}{\texttt{ndfa}}
\newcommand{\pda}{\texttt{pda}}
\newcommand{\tm}{\texttt{tm}}
\newcommand{\mttm}{\texttt{mttm}}
\newcommand{\sig}{\texttt{\(\Sigma\)}}

\newcommand{\delt}{\texttt{\(\delta\)}}

\newcommand{\quot}{\texttt{\textquotesingle{}}}

\newcommand{\step}{\texttt{$\vdash$}}

\newcommand{\arrow}{\(\rightarrow\)}

\newcommand{\flatt}{\texttt{FLAT}}
\newcommand{\gviz}{\texttt{Graphviz}}

\definecolor{darkgreen}{rgb}{0.01, 0.75, 0.24}
\definecolor{denim}{rgb}{0.08, 0.38, 0.74}
\definecolor{deepsaffron}{rgb}{1.0, 0.6, 0.2}
\definecolor{deeppink}{rgb}{1.0, 0.08, 0.58}
\definecolor{deeppeach}{rgb}{1.0, 0.8, 0.64}
\definecolor{dollarbill}{rgb}{0.52, 0.73, 0.4}
\definecolor{heliotrope}{rgb}{0.87, 0.45, 1.0}
\definecolor{cornflowerblue}{rgb}{0.39,0.58,0.93}
\definecolor{ForestGreen}{rgb}{0.13, 0.55, 0.13}

\title{Design Support for Multitape Turing Machines}

\author{Marco T. Moraz\'{a}n
\institute{Seton Hall University}
\email{morazanm@shu.edu}
\and
Oliwia Kempinski
\institute{University of Maryland}
\email{okempins@umd.edu}
\and
Andr\'{e}s M. Garced
\institute{Seton Hall University}
\email{maldona2@shu.edu}
}

\def\titlerunning{Design Support for Multitape Turing Machines}
\def\authorrunning{Moraz\'{a}n, Kempinski, and Garced}

\begin{document}

\maketitle

\begin{abstract}
Many Formal Languages and Automata Theory courses introduce students to Turing machine extensions. One of the most widely-used extensions endows Turing machines with multiple tapes. Although multitape Turing machines are an abstraction to simplify Turing machine design, students find them no less challenging. To aid students in understanding these machines, the \fsm{} programming language provides support for their definition and execution. This, however, has proven insufficient for many students to understand the operational semantics of such machines and to understand why such machines accept or reject a word. To address this problem, three visualization tools have been developed. The first is a dynamic visualization tool that simulates machine execution. The second is a static visualization tool that automatically renders a graphic for a multitape Turing machine's transition diagram. The third is a static visualization tool that automatically renders computation graphs for multitape Turing machines. This article presents these tools and illustrates how they are used to help students design and implement multitape Turing machines. In addition, empirical data is presented that suggests these tools are well-received and found useful by students.
\end{abstract}

\section{Introduction}

In many Formal Languages and Automata Theory (\flatt{}) courses, students are taught about Turing machine (\tm{}) extensions. These extensions are useful to illustrate the robustness of \tm{}s \cite{Sipser} and to simplify construction proofs (e.g., the equivalence of grammars and \tm{}s \cite{Lewis,PBFLAT,Rich}). One of the most widely discussed extensions in \flatt{} textbooks is multitape Turing machines (\mttm{}s). An \mttm{} is a Turing machine that has \texttt{n} tapes and \texttt{n} reading heads (one for each tape), where \texttt{n$\geq$1}. A standard \tm{}, for example, is an \mttm{} for which \texttt{n}=1. Computationally speaking, \mttm{}s and \tm{}s have the same power \cite{Lewis,PBFLAT,Rich}. For students, it is interesting to study \mttm{}s, because oftentimes it is easier to design and implement an \mttm{} rather than a \tm{}. 

Despite offering a higher-level of abstraction than standard \tm{}s, \mttm{}s remain difficult for first-time \flatt{} students to design. Design difficulties are accentuated by courses that have students design such machines by pencil and paper. In such courses, students are asked to design without being able to experiment nor get immediate feedback from a compiler or an interpreter. Such timely feedback has proven useful in other areas of Computer Science when exposing students for the first time to a body of work \cite{Race,Venables}. Challenges to understanding can also arise in courses that have students design such machines using visualization tools like \texttt{JFLAP} \cite{Rodger,RodgerII}. In our experience, for instance, students are distracted from machine design by trying to manually render an \mttm{}'s transition diagram in an appealing manner. In addition, visualizations require users to learn their interface and this can place an extraneous cognitive load on students \cite{Hegarty,Sweller}. To be effective and reduce such a load, visualizations need to have an easy-to-learn interface and provide representations that reflect or behave as the object itself \cite{Hutchins}.

To address the shortcomings outlined above, \fsm{}, a functional domain-specific language for the \flatt{} classroom, has been developed \cite{fsm,fsm-viz}. This language allows programmers, for example, to define, execute, and validate \mttm{}s. In addition to unit testing, immediate feedback is provided by a tailor-made error-messaging system that captures constructor misuse \cite{fsm-errors2,fsm-errors}. This language infrastructure allows students to experiment and debug their designs before submitting for grading or attempting a proof. This infrastructure, however, has proven insufficient for some students to understand \mttm{}s. For this reason, a dynamic visualization tool to observe machine execution is integrated into \fsm{}. When a given word is accepted by a given machine, this tool presents a graphic highlighting the content of every tape, the position of every head, the previous state, and the current state for every step of the computation. In this manner, the user can observe state transitions, head movements, and mutations on each tape. 

The dynamic visualization tool proved to be effective in the classroom and, therefore, its use is integrated into the textbook of instruction \cite[Chapter~17]{PBFLAT}. Despite its effectiveness, students expressed the need for visualization tools to observe an \mttm{}'s transition diagram and to understand why a given word is rejected by a given \mttm{}. This article presents two new static visualization tools to address these needs. The first renders a visually appealing graphic for a given \mttm{}'s transition diagram. This tool helps students visualize the design of an \mttm{} implementation. In addition, it helps students visually validate their design's implementation in a piecemeal manner as it progresses with what we call \emph{phase-based transition diagrams}. The second tool automatically generates a graphic, given an \mttm{} and a word, for a computation graph \cite{Gidayu}. A computation graph summarizes all the paths in the computation tree for the given \mttm{} and  the given word. This tool helps students understand why a word is rejected. 

The article is organized as follows. \Cref{fsm} presents a brief introduction to the \mttm{} design methodology used in the classroom and the dynamic visualization tool. \Cref{td} presents the tool to generate \mttm{} transition diagrams and how it may be used to support the design process. \Cref{comp-graphs-mttms} presents the tool to generate \mttm{} computation graphs and how it is used to explain why a word is rejected. \Cref{data} discusses empirical data collected from students in a \flatt{} course at Seton Hall University that uses the presented visualization tools. \Cref{rw} discusses and contrasts with related work. Finally, \Cref{concls} presents concluding remarks and directions for future work.

\section{A Brief Introduction to \mttm{}s in \fsm}
\label{fsm}

\subsection{Interface}

\fsm{} provides constructors for a variety of state machines including deterministic finite-state automata (\dfa{}s), nondeterministic finite state automata (\ndfa{}s), nondeterministic pushdown automata (\pda{}s), nondeterministic Turing machines (\tm{}s), and multitape Turing machines (\mttm{}s). The constructor for \mttm{}s has the following signature:
\begin{alltt}
     \textcolor{pakistangreen}{make-mttm}: K \sig{} S F \delt{} n Y \arrow{} mttm
\end{alltt}
The constructor's inputs denote:\\
\begin{tabular}{llll}
  \texttt{K}: the states & \sig{}: the input alphabet & \texttt{S}\(\in\)\texttt{K}: the start state & \texttt{F}\(\subseteq\)\texttt{K}: the final states \\ \\
  \texttt{\delt{}}: the transition relation & \texttt{n}: the number of tapes & \texttt{Y}\(\in\)\texttt{F}: the accept state\\ \\
\end{tabular}\\
Each transition rule in \delt{} has the following structure:
\begin{alltt}
     (list (list state (list symbol\(\sp{\texttt{n}}\))) (list state (list action\(\sp{\texttt{n}}\))))
\end{alltt}
The first tuple contains a state and a list of \texttt{n} alphabet elements read (one for each tape). The second tuple contains the state the machine transitions to and the list of actions taken by each of the \texttt{n} heads. An action either moves the head left, moves the head right, or mutates the tape by overwriting the read symbol with the given alphabet symbol in the rule. A transition may be used if the machine's current state matches the state in the first tuple and the elements read on the tapes (from 0 to \texttt{n}-1) match the list of symbols in the first tuple (from left to right). To illustrate a transition rule, consider:
\begin{lstlisting}[language=racket,mathescape=true,numbers=none]
     $\quot{}$((F (c _ b c)) $\quot{}$(C (R L _ d)))
\end{lstlisting}
This rule may be used when the machine is in state \texttt{F} reading a \texttt{c}, a blank, a \texttt{b}, and a \texttt{c}, respectively, on tapes 0--3. When used, the machine moves the head on tape 0 to the right, moves the head on tape 1 to the left, mutates the \texttt{b} read on tape 2 to a blank, mutates the \texttt{c} read on tape 3 to a \texttt{d}, and transitions to state \texttt{C}.

There is an observer to extract each component of an \mttm{}. For example, given an \mttm{}, \textcolor{pakistangreen}{\texttt{sm-states}} extracts the states and \textcolor{pakistangreen}{\texttt{sm-numtapes}} extracts the number of tapes. Machine application requires the input word on tape 0 and the initial head position on tape 0 (all other tapes are initially blank and their heads are on position 0). The observer \textcolor{pakistangreen}{\texttt{sm-apply}} applies the machine and returns the result of the computation\footnote{In this article, we are only concerned with \mttm{}s that decide or semidecide a language and not with those that compute the value of a function.}: \texttt{\quot{}accept} or \texttt{\quot{}reject}. The observer \textcolor{pakistangreen}{\texttt{sm-showtransitions}} applies the machine and returns a static trace of the configurations traversed on an accepting computation (i.e., one that reaches \texttt{Y}). If the machine rejects the given word then no trace is provided and only \texttt{\quot{}reject} is returned. Both of these observers may run forever if the \mttm{} only semidecides a language.

Three visualization primitives are integrated into the language: \textcolor{pakistangreen}{\texttt{(sm-visualize M)}}, \textcolor{pakistangreen}{\texttt{(sm-graph M)}}, and \textcolor{pakistangreen}{\texttt{(sm-cmpgraph M w [pos])}}. The first, given a machine, launches an \fsm{} dynamic visualization tool that lets the student step through a computation when given a word in the machine's language. The second generates a graphic for a given machine's transition diagram. The third, given a machine, an input word, and an optional initial head position on tape 0 (used only for \tm{}s and \mttm{}s), generates a computation graph that summarizes all computations possible when the word is rejected and, otherwise, summarizes a single accepting computation. The graphics for a transition diagram and a computation graph are generated using \gviz{} \cite{gviz1,gviz2}. This liberates \fsm{} programmers from having to manually render diagrams in an appealing manner. The extensions of these visualization tools for \mttm{}s and their classroom use motivates the work presented in this article.

\subsection{Design Process}

\begin{figure}[t!]
\begin{enumerate}
  \item Name the machine, specify the alphabet, and formulate the precondition.
  \item Write unit tests.
  \item Identify conditions that must be tracked as input is consumed, associate a state with each condition, and determine the start and final states.
  \item Formulate the transition relation.
  \item Implement the machine.
  \item Run the tests and, if necessary, redesign.
  \item Design, implement, and test an invariant predicate for each state.
  \item Prove L = L(M).
\end{enumerate}
\caption{The design recipe for multitape Turing machines.} \label{sm-dr}
\end{figure}

Machine development follows the methodology put forth by Moraz\'{a}n using a design recipe for state machines \cite{PBFLAT}. A design recipe, first introduced by Felleisen et al. to instruct beginners in programming \cite{HtDP2} and later expanded by Moraz\'{a}n to a 2-semester introduction for beginners \cite{drecipe,APS,APD}, consists of a systematic collection of steps that guide the implementation process. Each step has a concrete outcome, thus, creating a context in which students and instructors can discuss state machines. The design recipe for state machines adapted for \mttm{}s is displayed in \Cref{sm-dr}. It is noteworthy that it includes verification steps (i.e., steps 7 and 8), which are appropriate in a course that formally explores theoretical Computer Science. 

The steps may be described as follows for an \mttm{}, \texttt{M}, that decides or semidecides \texttt{L}:
\begin{description}[leftmargin=!,labelwidth=\widthof{\bfseries Step 8},labelindent=\parindent]
  \item[Step 1] Pick a descriptive machine name and specify the input alphabet. In addition, specify the machine's initial configuration (i.e., tape 0's value and head position).
  
  \item[Step 2] Write a thorough test suite. If the \mttm{} decides a language then this suite must include tests for words in and not in the machine's language. Otherwise, only tests for words in the machine's language must be included. Tests are written using \rackunit{} (e.g., \texttt{check-equal?}) \cite{RackUnit} and \textcolor{pakistangreen}{\texttt{sm-apply}} or \textcolor{pakistangreen}{\texttt{sm-showtransitions}}.
      
  \item[Step 3] Outline a design idea identifying conditions that must be tracked. These conditions define the role of states and for each a state name must be chosen. In addition, identify the starting and the final states.
      
  \item[Step 4] Develop the transition relation. For each transition, it is assumed that the role of the machine's current state is satisfied. The actions taken by the transition must guarantee the destination state's role is satisfied.
      
  \item[Step 5] Implement the machine using \textcolor{pakistangreen}{\texttt{make-mttm}} and the results of the previous steps.
  
  \item[Step 6] Run the tests and, if errors occur, perform guided debugging revisiting the previous steps.
  
  \item[Step 7] Write an invariant predicate for each state that determines if its role (from step 3) is satisfied. Each predicate takes as input a list of tape configurations. Each tape configuration contains the head's position and the tape's value.
      
  \item[Step 8] Argue machine correctness in two steps. The first proves by induction on the number of steps performed that the invariant predicates hold on accepting computations. The second, assuming invariant predicates always hold, prove that \texttt{L} = L(\texttt{M}).
\end{description}

\subsection{Illustrative Example}

\begin{figure}[t!]
\begin{lstlisting}[language=racket,mathescape=true]
#lang fsm
;; PRE: t0=(@ _ w), t0h=1, tapes 1-3 are empty, t1h-t3h=0
;; S precondition holds, starting state
;; C t0=(@ _ w), t0h=n<=|w|+3, 
;;    as, bs, and cs in t0[1..n-1] copied, respectively, to t1, t2, and t3
;; D t0=(@ _ w), t0h=n<=|w|+2, t0[t0h]=a, 
;;    as, bs, and cs in t0[1..n] copied, respectively, to t1, t2, and t3 
;; E t0=(@ _ w), t0h=n<=|w|+2, t0[t0h]=b, 
;;    as, bs, and cs in t0[1..n] copied, respectively, to t1, t2, and t3 
;; F t0=(@ _ w), t0h=n<=|w|+2, t0[t0h]=c, 
;;    as, bs, and cs in t0[1..n] copied, respectively, to t1, t2, and t3 
;; G t0=(@ _ w _), Let r=number as in t0, s=number bs in t0, t=number cs in t0,
;;                         t1=(_ a$^{\textcolor{darkorange}{r}}$ _), t2=(_ b$^{\textcolor{darkorange}{s}}$ _), t3=(_ c$^{\textcolor{darkorange}{t}}$ _), t1h<=r, t2h<=s, t3h<=t, 
;;                         |[t1h..r+1]|=|[t2h..s+1]|=|[t3h..t+1]|
;; Y t0=(@ _ w _), Let r=number as in t0, s=number bs in t0, t=number cs in t0, r=s=t, 
;;    final and accepting state
(define EQABC
  (make-mttm $\quot{}$(S Y C D E F G) $\quot{}$(a b c) $\quot{}$S $\quot{}$(Y)
             '(((S (_ _ _ _)) (C (R R R R)))
               ((C (a _ _ _)) (D (a a _ _)))
               ((D (a a _ _)) (C (R R _ _)))
               ((C (b _ _ _)) (E (b _ b _)))
               ((E (b _ b _)) (C (R _ R _)))
               ((C (c _ _ _)) (F (c _ _ c)))
               ((F (c _ _ c)) (C (R _ _ R)))
               ((C (_ _ _ _)) (G (_ L L L)))
               ((G (_ _ _ _)) (Y (_ _ _ _)))
               ((G (_ a b c)) (G (_ L L L))))
             4 $\quot{}$Y))       
(check-equal? (sm-apply EQABC $\quot$(@ _ a a b b a c c) 1)     $\quot{}$reject)
(check-equal? (sm-apply EQABC $\quot$(@ _ a a a) 1)             $\quot{}$reject)
(check-equal? (sm-apply EQABC $\quot$(@ _ c c a b b) 1)         $\quot{}$reject)
(check-equal? (sm-apply EQABC $\quot$(@ _) 1)                   $\quot{}$accept)
(check-equal? (sm-apply EQABC $\quot$(@ _ a c c b a b) 1)       $\quot{}$accept)
(check-equal? (sm-apply EQABC $\quot$(@ _ c c a b a b a b c) 1) $\quot{}$accept)
\end{lstlisting}
\caption{An \mttm{} to decide  \texttt{L}=\{\texttt{w} $|$ \texttt{w} has equal number of \texttt{a}s, \texttt{b}s, and \texttt{c}s\}.} \label{mttm-eqabc}
\end{figure}

Consider designing and implementing a machine for \texttt{L}=\{\texttt{w} $|$ \texttt{w} has equal number of \texttt{a}s, \texttt{b}s, and \texttt{c}s\}. \Cref{mttm-eqabc} displays a 4-tape \mttm{} implementation obtained from following the steps of the design recipe. The name, alphabet, and precondition to satisfy Step 1 are displayed on lines 17, 18, and 2\footnote{The tape's left-end marker is denoted by \texttt{@} and the machine upon reading it must move the head right.}, where \texttt{t0}--\texttt{t3} denote the tapes and \texttt{t0h}--\texttt{t3h} denote the respective head positions. The precondition states that \texttt{t0}'s tape contains \texttt{w} preceded by a blank, \texttt{t0}'s head is on this blank, and all other tapes are empty with their heads on the initial blank. The test suite to satisfy Step 2 is displayed on lines 30--35. 

To satisfy Step 3, the machine operates in three phases as follows:
\begin{enumerate}
  \item In the starting state, \texttt{S}, the machine reads a blank on all tapes, moves all heads right, and transitions to state \texttt{C} (the control state for phase two). 
      
  \item In this phase, the machine repeatedly moves from \texttt{C} to either \texttt{D}, \texttt{E}, or \texttt{F} to copy an \texttt{a}, \texttt{b}, or \texttt{c} to, respectively, tape 1, tape 2, or tape 3. Once the copying is completed, the heads are moved right and the machine loops back to \texttt{C} to copy, if any, the remaining word elements. To start the next phase, the machine transitions from \texttt{C} to \texttt{G} when it reads a blank on all tapes and moves the heads on tapes 1--3 to the left.
      
  \item In this phase, the machine repeatedly moves the heads on tapes 1--3 left as long as an \texttt{a}, \texttt{b}, and \texttt{c} are, respectively, read on tapes 1--3. Eventually, if blanks are read on tapes 1--3 the machine moves to \texttt{Y} and accepts.
\end{enumerate}
The states are documented on lines 3--16. State \texttt{S} is identified as the starting state and state \texttt{Y} is identified as the only final state as well as the accepting state. 
Briefly, consider the documentation for state \texttt{G}. It states that \texttt{t0} contains \texttt{w} between blanks, that \texttt{t0}'s \texttt{a}s, \texttt{b}s, and \texttt{c}s are copied, respectively, to \texttt{t1}, \texttt{t2}, and \texttt{t3}, and that there are matching \texttt{a}s, \texttt{bs}, and \texttt{c}s on tapes \texttt{t1}--\texttt{t3} after their respective heads. We trust that this suffices to parse the descriptions of the other states. For this step, as the reader can appreciate, the descriptions are informal. We note, however, that for each state, a description is implemented as a predicate to satisfy Step 7.

The answer to satisfy Step 4 is displayed on lines 19--28. Each transition is designed considering the state descriptions developed for the previous step. For instance, in state \texttt{C} all the \texttt{a}s, \texttt{b}s, and \texttt{c}s  on \texttt{t0} before \texttt{t0}'s head are copied, respectively, to \texttt{t1}--\texttt{t3}. If an \texttt{a} is read, then it must be copied to \texttt{t1} without mutating \texttt{t0}, which leaves the machine in a state that satisfies state \texttt{D}'s role. Therefore, a needed transition is: \texttt{(list \quot{}(C (a \_ \_ \_))  \quot{}(D (a a \_ \_)))}. A similar analysis is performed to develop the remainder of the transitions.

Step 5 is satisfied by the code displayed in \Cref{mttm-eqabc}. Step 6 is satisfied by running the tests and verifying that no errors are thrown and no tests fail. Steps 7 and 8 are not directly relevant to understand the rest of the article. In the interest of brevity, we omit describing these steps. The interested reader is referred to previously published work \cite{PBFLAT}.

\subsection{Dynamic Visualization}
\label{dynamic-viz}

This (previously implemented) visualization tool allows the user to specify the input word and the initial head position on tape 0. Once these are specified, the user may move forward and backwards in the computation one step at a time. At each step, the user may see the following features:
\begin{itemize}
  \item[$\circ$] All tapes horizontally rendered and vertically aligned by tape positions
  
  \item[$\circ$] The head position on each tape, highlighting in color the element under the head
  
  \item[$\circ$] The numbering of positions on each tape
  
  \item[$\circ$] The transition rule used in the last step
  
  \item[$\circ$] The current and previous states
  
  \item[$\circ$] If invariant state predicates are provided, then the current state is rendered in a color box: green if the invariant holds and red if the invariant does not hold
\end{itemize}
If the number of tapes does not fit in the visualization's frame, the user may scroll vertically to observe a different subset of tapes. Similarly, if the content of any tape does not fit in the visualization's frame, the user may scroll horizontally to observe different tape positions. We note that this visualization strategy differs from that used for other state machines, which allow stepping through a computation using the machine's transition diagram.

\begin{figure}[t!]
\centering
\includegraphics[scale=1.5]{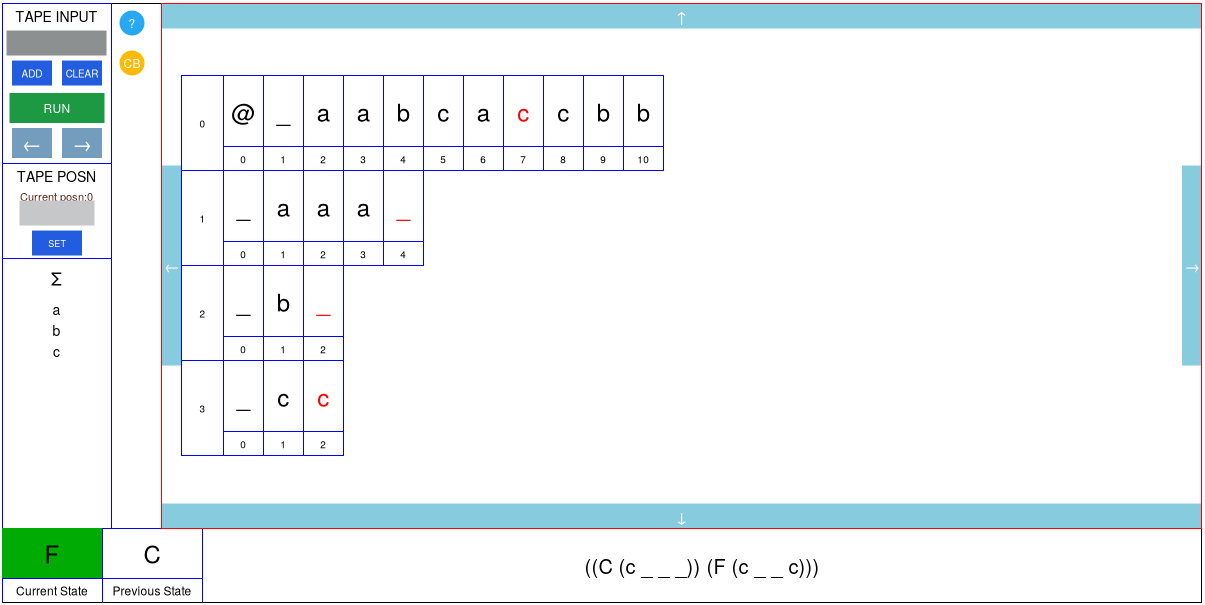}
\caption{A snapshot of the dynamic visualization tool for \mttm{}s.} \label{mttm-viz}
\end{figure}

\Cref{mttm-viz} displays a snapshot of the tool. In the left column (excluding the bottom row), the user can enter a word (or clear it) for tape 0 and set this tape's initial head position. The run button is used to run the computation and the arrow buttons are used to move the visualization backwards and forward. Finally, the machine's alphabet is displayed. The main visualization graphic is immediately to the right. The content of each tape is displayed along with the index of each element. The element under each head is highlighted in red to indicate the head position. In \Cref{mttm-viz}, the machine is reading the second \texttt{c} in the input word (on tape 0) and has copied the previous elements to tapes 1--3. The heads on tapes 0--3 are, respectively, at positions 7, 4, 2, and 2. Finally, at the bottom, the last transition rule used is displayed along with, \texttt{F} and \texttt{C}, the current and previous states. This provides the context to understand the state of the tapes. The machine has moved from state \texttt{C} to state \texttt{F} reading a \texttt{c} on tape 0 and copying it tape 3 using mutation. Finally, state \texttt{F} is rendered in a green box indicating that the state invariant predicate provided for it holds. Thus, affirming \texttt{F}'s role as defined in Step 3 of the design recipe for state machines (assuming, of course, that said invariant is correctly implemented).

The tool is effective for explaining why a given word is accepted by a given machine. As pointed out by students, the tool falls short of providing a visual representation of the machine's structure (i.e., its transition diagram) and of providing any help to understand why a word is rejected. The former is a problem, because it is not always easy to mentally visualize how the machine ought to operate by reading the textual transition relation. The latter is a problem, because it leaves students wondering why, for example, a word that ought to be accepted is rejected. This contrasts sharply with the support provided for other machine types in \fsm{} for which transition diagrams and computation graphs \cite{fsm-cmpgraphs} are generated. These shortcomings motivate the rest of the work presented in this article on generating transition diagrams and computation graphs for \mttm{}s.

\section{Transition Diagrams for \mttm{}s}
\label{td}

Graph drawing is a complex and costly task that requires automation \cite{kaufmann} to facilitate the visual analysis of various kinds of connected systems \cite{Kruja}, such as an \mttm{}'s transition relation.  Without automation, \fsm{} programmers are distracted from their primary task which is, for example, to design and implement an \mttm{}\footnote{There is evidence that drawing can improve memory retention and recall \cite{Roberts,Wammes}. That is, the act of preparing to draw or the act of drawing an item makes the item more readily retrievable. The item must exist, concretely or abstractly, before it is drawn. This, of course, is not the case when designing a new \mttm{}.}. In \fsm{}, such images are generated using \gviz{} \cite{gviz1,gviz2}. \gviz{} is an open-source software project used to help visualize information as abstract graphs and networks. An attractive \gviz{} feature is that nodes are automatically arranged in a visually appealing manner, thus, liberating users from having to perform this task. 

This section presents an extension to \textcolor{pakistangreen}{\texttt{sm-graph}} to generate transition diagrams for \mttm{}s. First, it summarizes the design idea behind their generation. Second, an illustrative example is presented.

\subsection{Design Idea}
\label{DI}

Three state varieties are defined:
\begin{itemize}
\item[$\circ$] start state 
\item[$\circ$] final state
  \begin{itemize}
    \item[$\ast$] rejecting final state 
    \item[$\ast$] accepting final state
  \end{itemize}
\item[$\circ$] ordinary state 
\end{itemize}
Each state is rendered with its name. A start state is rendered in green to distinguish it from all other states. There are two varieties of final states. The first, a rejecting final state, is any final state that is not the accepting state and is rendered as double circle. The second, the accepting state, is rendered as a double octagon. Finally, any state that is not a starting state nor a final state is coined an ordinary state and is rendered as a black circle.

A transition diagram only has labeled directed edges. Each denotes one or more machine transitions. For every transition denoted by an edge, there is a label on the edge separated by commas and, if necessary, stacked to prevent labels from becoming excessively long. A label consists of two lists. The first contains all the elements read on the \texttt{n} tapes: tape 0 to tape \texttt{n}-1 from left to right. The second list contains the actions taken by the \texttt{n} heads: tape 0 to tape \texttt{n}-1 from left to right. All transitions are rendered as solid black arrows. For example, consider the following transition rule:
\begin{alltt}
     ((C (a b _ a)) (D (R a L _)))
\end{alltt}
This rule states that if the machine is in state \texttt{C}, and an \texttt{a} is read on tape 0, a \texttt{b} is read on tape 1, a blank is read on tape 2, and an \texttt{a} is read on tape 3, then the machine transitions to state \texttt{D}, moves tape 0's head right, mutates the \texttt{b} read on tape 1 to an \texttt{a}, moves the head on tape 2 left, and mutates the \texttt{a} read on tape 3 to a blank. The edge for this rule is rendered as follows:
\begin{center}
\includegraphics[scale=0.4]{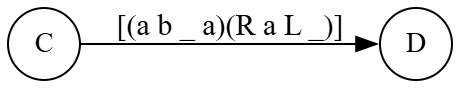}
\end{center}

\subsection{An Illustrative Example}
\label{transdiagram-ex}

\begin{figure}[t!]
\includegraphics[scale=0.35]{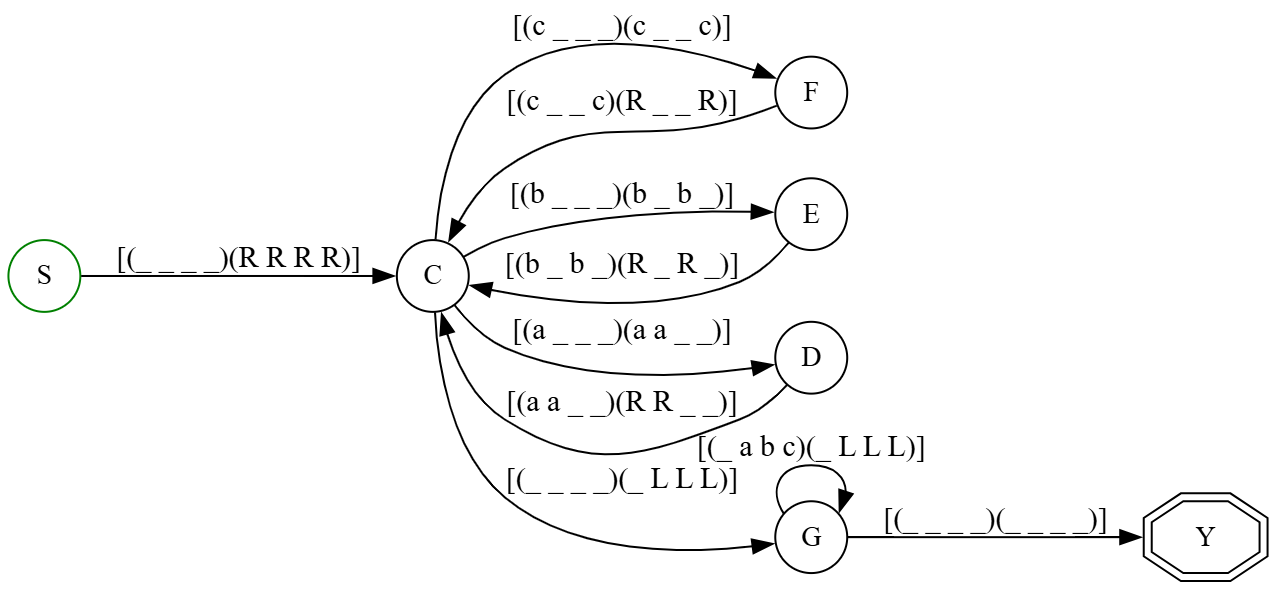}
\centering
\caption{Transition diagram for the \mttm{} from \Cref{mttm-eqabc}.} \label{mttm-eqabc-td}
\end{figure}

The transition diagram for \texttt{EQABC} from \Cref{mttm-eqabc} is displayed in \Cref{mttm-eqabc-td}. From the graphic, the starting state, \texttt{S}, and, \texttt{Y}, the accepting state are immediately discernable. In addition, the relationship between the states is easier to discern by looking at the graphic than by reading the textual transition relation in \Cref{mttm-eqabc}. Also more easily discernable are the three phases the machine operates in. The first phase is implemented by the edge from \texttt{S} to \texttt{C}. The second phase is implemented by the loops from \texttt{C} to, respectively, \texttt{D}, \texttt{E}, and \texttt{F}. The third phase is implemented by the edge from \texttt{C} to \texttt{G}, the loop on \texttt{G}, and the edge from \texttt{G} to \texttt{Y}. The visually appealing transition diagram also allows both students and instructors to easily distinguish the nuances of the operational semantics of the machine (i.e., the phases). For instance, it is straightforward to observe that the loop between \texttt{C} and \texttt{F} copies a \texttt{c} from tape 0 to tape 3. Similarly, it is also straightforward to observe that in state \texttt{G} the heads on tapes 1--3 are moved left as long as an \texttt{a}, a \texttt{b}, and a \texttt{c} are matched on the respective tapes. Clearly, the same nuances may be ascertained from the program in \Cref{mttm-eqabc}, but requires more effort given that the transition rules must be mentally parsed. Automatic rendering of transition diagrams allows students, in many cases, to quickly validate their design visually.

\begin{figure}[t!]
\captionsetup[subfigure]{justification=centering}
\centering
\begin{subfigure}[b]{0.35\textwidth}
  \centering
  \includegraphics[scale=0.3]{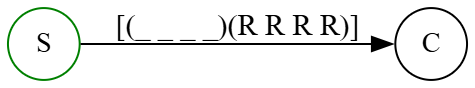}
  \caption{Phase 1.} \label{p1-eqabc}
\end{subfigure}
\qquad
\begin{subfigure}[b]{0.55\textwidth}
  \centering
  \includegraphics[scale=0.3]{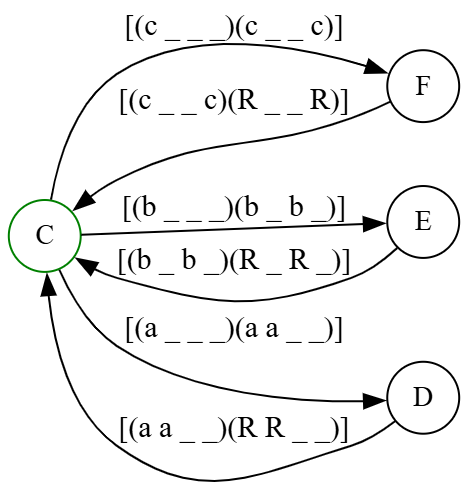}
  \caption{Phase 2.} \label{p2-eqabc}
\end{subfigure}

\begin{subfigure}[b]{\textwidth}
  \centering
  \includegraphics[scale=0.3]{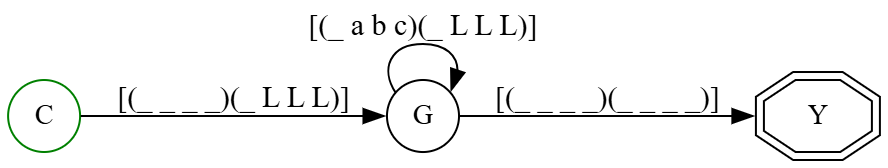}
  \caption{Phase 3.} \label{p3-eqabc}
\end{subfigure}
\caption{Phase-based transition diagrams for \texttt{EQABC}.} \label{phase-tds}
\end{figure}

Despite the visually appealing layout of an \mttm{}'s transition diagram, visually validating a complete design is a difficult and an error-prone process for large machines. In such cases, it is helpful to use what we coin \emph{phase-based transition diagrams}. In a phase-based transition diagram, only the subset of states and the subset of transitions provided to the constructor are displayed (presumably relevant to a phase). Such transition diagrams may be generated as transitions are added piecemeal by phases to the transition relation's implementation or by commenting out edges and/or nodes after the transition relation is entirely coded. \Cref{phase-tds} displays the phase-based transition diagrams for \texttt{EQABC} from \Cref{mttm-eqabc}. Each allows for the nuances of each phase to be more easily observed and, as a consequence, for each phase to be more easily debugged when necessary. For instance, as displayed in \Crefrange{p1-eqabc}{p3-eqabc}, the starting state for each phase may be highlighted in green by providing it as the starting state to the constructor and it is clear that the loop between \texttt{C} and \texttt{F} first copies a \texttt{c} from tape 0 to tape 3 and then moves the heads on tape 0 and tape 3 to the right.

\section{\fsm{} Computation Graphs for \mttm{}s}
\label{comp-graphs-mttms}

In general, the \mttm{} dynamic visualization tool helps students understand why a word is accepted and is preferred over having to mentally parse the output provided by \textcolor{pakistangreen}{\texttt{sm-showtransitions}}. As observed earlier, however, a significant number of students feel at a loss when an \mttm{} rejects a word. This has motivated our work on \fsm{} computation graphs for \mttm{}s. An \fsm{} computation graph is a finite static visualization to help explain nondeterministic behavior. When a word is rejected, it summarizes all the paths in the computation tree. When a word is accepted, it summarizes a single path for an accepting computation. This work extends \textcolor{pakistangreen}{\texttt{sm-cmpgraph}} in a manner that is consistent with what is done for other types of state machines.

In an \mttm{} computation graph, nodes represent states, not machine configurations\footnote{In \texttt{Gidayu}, for example, nodes in computation graphs for finite-state machines and pushdown automata represent machine configurations \cite{Gidayu}.}, and directed edges represent transitions. Such graphs only contain the states and transitions traversed by any computation performed on a given word. As a consequence, an \mttm{} computation graph is always a subgraph of the given \mttm{}’s transition diagram, which allows an \fsm{} programmer to easily see the connection between the machine’s transition diagram and the visualization provided. The states in which any computation terminates are highlighted in crimson. Thus, allowing the user to visually ascertain if there are terminating computations and if the given word is accepted or rejected. If a final state is highlighted in crimson, then there is at least one computation that terminates. If the accepting state is part of an \mttm{} computation graph then it is highlighted in crimson (no transitions are allowed after reaching a final state) and serves as visual confirmation that the given word is accepted. Given that all that is needed to establish that the given word is in the given machine's language is an accepting computation, \mttm{} computation graphs are trimmed to only contain the states and the edges traversed by a single accepting computation. Given that all computations must reject to establish that the given word is not in the given machine's language, no such trimming is performed when an accepting computation is not found.

It is well-known that, like standard Turing machines, \mttm{}s may only semidecide a language and, therefore, can perform computations that do not terminate. This means that it is impossible to conclusively determine that a given word is not in the given machine's language (e.g., word membership in a \tm{}'s language is undecidable \cite{Lewis,PBFLAT,Rich}). Therefore, it is not always possible to construct a computation graph for an \mttm{}. For such cases, the best we can do is construct a partial computation graph that we refer to as an \emph{approximation computation graph}. An approximation computation graph is constructed using a threshold parameter for the maximum number of steps a computation may perform, where a step is the application of a transition. If the number steps in a computation reaches this threshold then it is cutoff and, in essence, treated as if it has terminated. In an approximation computation graph, the state in which a computation is cut off is highlighted in gold. In this manner, the \fsm{} programmer is made aware that the computation graph is an approximation\footnote{To alleviate the cutoff threshold being too small or too large, users may customize its value.}. 

The next subsections describe how the work done on computation graphs for finite-state machines, pushdown automata, and Turing machines \cite{fsm-cmpgraphs} is extended to \mttm{}s. We first outline the design idea behind the generation of \mttm{} computation graphs. After that, two illustrative examples are presented: one for a deterministic \mttm{} and one for a nondeterministic \mttm{}.

\subsection{Design Idea for \mttm{} Computation Graphs}

In order to correctly apply node coloring, four varieties of \mttm{} computation graph edges are defined:
\begin{description}[leftmargin=!,labelwidth=\widthof{\bfseries cutoff special edge},labelindent=\parindent]
\item[regular edge] Represents a computation step that is not cut off, does not lead to a final state, and is not the last step of a computation. The state transitioned into requires no color highlighting. We denote such an edge as an \texttt{r-edge}.
\item[special edge] Represents a computation step that is not cut off, and either leads to a final state or is the last step of a computation. The destination state's outline is highlighted in crimson. We denote such an edge as a \texttt{sp-edge}.
\item[cutoff edge] Represents a computation step that is cut off, does not lead to a final state, and is not the last step of a computation. The destination state's filling is highlighted in gold. We denote such an edge as a \texttt{co-edge}
\item[cutoff special edge] Represents a step in at least two different computations. In one computation it is a \texttt{co-edge} and in another it is a \texttt{sp-edge}. The destination state's filling is highlighted in gold for the former and its outline is highlighted in crimson for the latter. We denote such an edge as a \texttt{cosp-edge}
\end{description}
A transition, \texttt{t}, may be used by multiple computations performed by a nondeterministic \mttm{}. It is possible that for some computations the use of \texttt{t} is represented as a \texttt{r-edge} while in other computations it is represented as a \texttt{sp-edge} or a \texttt{co-edge}. For \mttm{} computation graph rendering, edges requiring highlighting take precedence over edges that do not require highlighting. Thus, if \texttt{t} is represented as an \texttt{r-edge} in one computation and as a \texttt{co-edge} in another computation, then the \texttt{co-edge} is used to render the computation graph. When \texttt{t} is represented as a \texttt{sp-edge} and as a \texttt{co-edge} in two different computations, a \texttt{cosp-edge} is used to properly render the computation graph by having the destination state highlighted in both crimson and gold.

To lower the extraneous cognitive load, every computation graph generated contains an informative message. The message either indicates that the word is accepted/rejected, or that there are cut off computations. Although this information can be extrapolated from the graphic, students find such messages useful when they first use \mttm{} computation graphs.

\subsection{An Illustrative Example Using a Deterministic \mttm{}}
\label{d-mttm-cg-ex}

\begin{figure}[t!]
\captionsetup[subfigure]{justification=centering}
\centering
\begin{subfigure}[b]{\textwidth}
  \centering
  \includegraphics[scale=0.52]{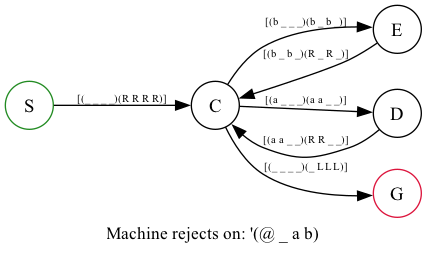}
  \caption{Computation graph for \quot\texttt{(@ \textunderscore{} a b)}.} \label{Mviz8}
\end{subfigure}
\hfill
\begin{subfigure}[b]{\textwidth}
  \centering
  \includegraphics[scale=0.52]{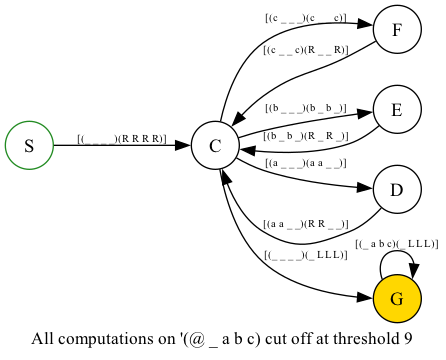}
  \caption{Computation graph for \quot\texttt{(@ \textunderscore{} a b c)} and threshold 9.} \label{Mviz5}
\end{subfigure}
\hfill
\begin{subfigure}[b]{\textwidth}
  \centering
  \includegraphics[scale=0.52]{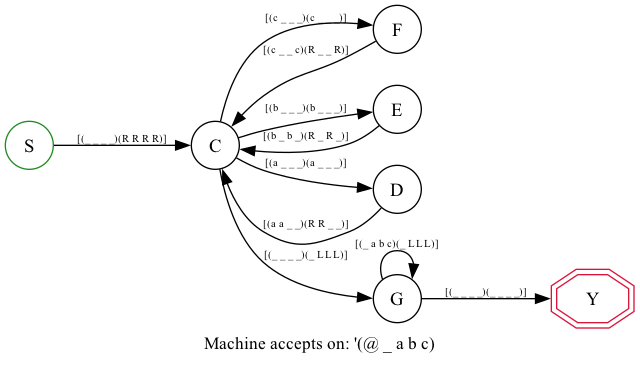}
  \caption{Computation graph for \quot\texttt{(@ \textunderscore{} a b c)} and no threshold.} \label{Mviz6}
\end{subfigure}
\caption{Computation graphs for \texttt{EQABC} on reject, cutoff, and accept.} \label{Mviz7}
\end{figure}

To illustrate \mttm{} computation graphs, we use \texttt{EQABC} from \Cref{mttm-eqabc}. First, consider applying \texttt{EQABC} to $\quot{}$(@ \textunderscore{} a b). Its \mttm{} computation graph is displayed in \Cref{Mviz8}. It is immediately clear that the machine rejects because the only crimson state, \texttt{G}, is not, \texttt{Y}, the accepting state. Observe that state \texttt{F} is not part of the computation graph. This means that there is no computation that traverses \texttt{F}. Therefore, to explain why \texttt{EQABC} rejects it is not necessary to consider \texttt{F}. Rejection is explained by reconstructing computations from the starting configuration and only using the transitions contained in the computation graph as follows\footnote{Each configuration contains a state, the head positions, from left to right, for tapes 0--3, and the respective contents of each tape.}:
\begin{alltt}
     (S (1 0 0 0) ((@ _ a b)(_)(_)(_))) 
       \step (C (2 1 1 1) ((@ _ a b)    (_ _)    (_ _)    (_ _)))
       \step (D (2 1 1 1) ((@ _ a b)    (_ a)    (_ _)    (_ _)))
       \step (C (3 2 1 1) ((@ _ a b)    (_ a _)  (_ _)    (_ _)))
       \step (D (3 2 1 1) ((@ _ a b)    (_ a _)  (_ b)    (_ _)))
       \step (C (4 2 2 1) ((@ _ a b _)  (_ a _)  (_ b _)  (_ _)))
       \step (G (4 1 1 0) ((@ _ a b _)  (_ a _)  (_ b _)  (_ _)))                   
\end{alltt}
In last configuration, the machine is in state \texttt{G}, a blank is read on tape 0, an \texttt{a} is read on tape 1, a \texttt{b} is read on tape 2, and a blank is read on tape 3. Given that \texttt{EQABC} has no applicable transition, the machine halts, thus, explaining why \texttt{G} is highlighted in crimson. The informative message, in this case, indicates that the word is rejected.

\Cref{Mviz5} displays the computation graph obtained by applying \texttt{EQABC} to $\quot{}$(@ \textunderscore{} a b c) using a cutoff threshold of 9 steps\footnote{The threshold has been set low for this example to illustrate cutoff behavior.}. From the computation graph, we can immediately discern that no accepting computation was found given that, \texttt{Y}, the accepting state is not contained in it and that at least one computation is cut off given that state \texttt{G} is highlighted in gold. In this example, there is also a single computation:
\begin{alltt}
     (S (1 0 0 0) ((@ _ a b c)(_)(_)(_))) 
       \step (C (2 1 1 1)  ((@ _ a b c)    (_ _)   (_ _)   (_ _)))
       \step (D (2 1 1 1)  ((@ _ a b c)    (_ a)   (_ _)   (_ _)))
       \step (C (3 2 1 1)  ((@ _ a b c)    (_ a _) (_ _)   (_ _)))
       \step (E (3 2 1 1)  ((@ _ a b c)    (_ a _) (_ b)   (_ _)))
       \step (C (4 2 2 1)  ((@ _ a b c)  (_ a _) (_ b _) (_ _)))
       \step (F (4 2 2 1)  ((@ _ a b c)    (_ a _) (_ b _) (_ c)))
       \step (C (5 2 2 2)  ((@ _ a b c _)  (_ a _) (_ b _) (_ c _)))
       \step (G (5 1 1 1)  ((@ _ a b c _)  (_ a _) (_ b _) (_ c _)))
       \step (G (5 0 0 0)  ((@ _ a b c _)  (_ a _) (_ b _) (_ c _)))                                        
\end{alltt}
The computation copies each element of the given word to the auxiliary tapes in 9 steps to end in state \texttt{G}. At this point, the cutoff is triggered and the computation is halted. This is why \texttt{G} is highlighted in gold. The informative message, in this case, indicates the cutoff threshold.

The \fsm{} programmer may raise the cutoff threshold to try to confirm, for example, that the machine is not caught in an infinite loop in state \texttt{G}. Upon doing so, the \mttm{} computation graph obtained is displayed in \Cref{Mviz6}. It clearly communicates that the word is accepted, given that it contains, \texttt{Y}, the accepting state. To accept, the machine extends the computation above with one step: 
\begin{alltt}
       \step (Y (5 0 0 0) ((@ _ a b c _)(_ a _)(_ b _)(_ c _)))             
\end{alltt}
The informative message, in this case, indicates that the word is accepted.

\subsection{An Illustrative Example Using a Nondeterministic \mttm{}}
\label{nd-mttm-cg-ex}

\begin{figure}[t!]
\captionsetup[subfigure]{justification=centering}
\centering
\includegraphics[scale=0.50]{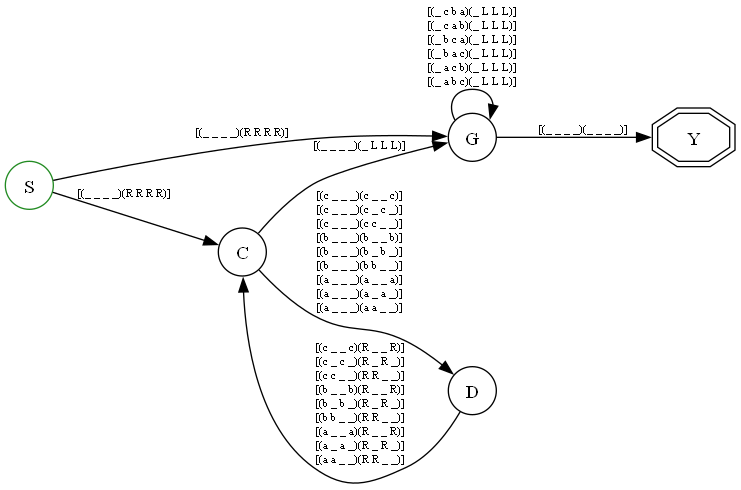}
\caption{Nondeterministic \mttm{} transition diagram for \texttt{L}=\{\texttt{w} $|$ \texttt{w} has equal number of \texttt{a}s, \texttt{b}s, and \texttt{c}s\}.} \label{eqabc-nd-td}
\end{figure}

To illustrate computation graphs for nondeterministic \mttm{}s, consider, \texttt{EQABC-ND}, a different 4-tape machine design for \texttt{L}=\{\texttt{w} $|$ \texttt{w} has equal number of \texttt{a}s, \texttt{b}s, and \texttt{c}s\} that operates in three phases as follows:
\begin{enumerate}
  \item Blanks read on all tapes, all heads move right, and transition to the phase 2 control state. 
      
  \item The machine nondeterministically copies the input word elements to the auxiliary tapes.
      
  \item The machine repeatedly moves the heads on tapes 1--3 left as long as an \texttt{a}, a \texttt{b}, and a \texttt{c} are read on the auxiliary tapes. If blanks are read on the auxiliary tapes then the machine moves to accept.
\end{enumerate}
The transition diagram for this design is displayed in \Cref{eqabc-nd-td}\footnote{For interested readers, the \fsm{} code is found in \Cref{A1}, albeit without outlining the steps of the design recipe.}. Phase 1 is implemented by the transition from \texttt{S} to \texttt{C}. Phase 2 is implement by the loop between \texttt{C} and \texttt{D}. In this loop, the machine nondeterministically copies the read element to one of the auxiliary tapes and returns to \texttt{C} to copy, if any, the next element. Phase 3 is implemented by the remaining transitions.

\begin{figure}[t!]
\captionsetup[subfigure]{justification=centering}
\centering
\begin{subfigure}[b]{0.44\textwidth}
  \includegraphics[scale=0.4]{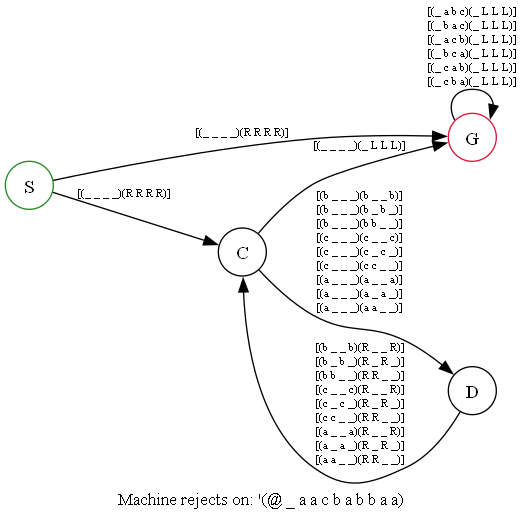}
  \caption{On reject for \quot{}\texttt{(\_ a a c b a b b a a)}.} \label{eqabc-nd-reject}
\end{subfigure}
\qquad
\begin{subfigure}[b]{0.47\textwidth}
  \includegraphics[scale=0.4]{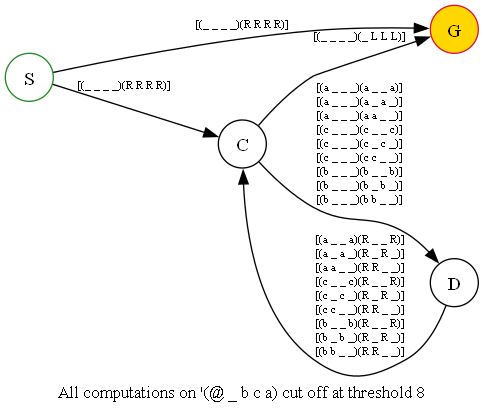}
  \caption{On cut off of 8 for \quot{}\texttt{(\_ b c a)}.} \label{eqabc-nd-cutoff}
\end{subfigure}
\hfill
\begin{subfigure}[b]{\textwidth}
  \centering
  \includegraphics[scale=0.4]{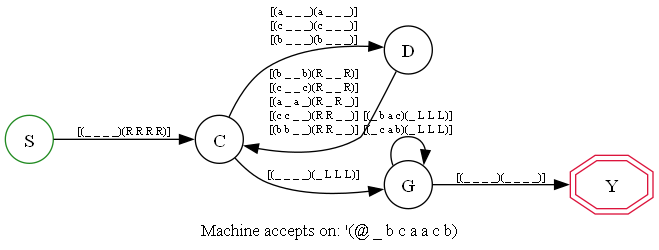}
  \caption{Computation graph on accept for \quot{}\texttt{(\_ b c a a c b)}.} \label{eqabc-nd-accept}
\end{subfigure}
\caption{Computation graphs for \texttt{EQABC-ND} on reject, cutoff, and accept.} \label{eqabc-nd-cmpgraphs}
\end{figure}

Consider applying \texttt{EQABC-ND} to  \quot{}\texttt{(\_ a a c b a b b a a)}. The computation graph for this application is displayed in \Cref{eqabc-nd-reject}. We can immediately observe that the only state highlighted in crimson, \texttt{G}, is not the accepting state and there are no computations cut off. Therefore, the machine rejects. To understand why the machine rejects, it is easiest to reason about the phases of the machines using the \mttm{} computation graph as an aid. There is a computation in which the machine nondeterministically moves to \texttt{G} in phase 1, where no transition rules apply and the computation halts. Thus, explaining why the state is highlighted in crimson. There are also computations in which the machine nondeterministically moves to state \texttt{C} in phase 1 and, in phase 2, copies all the elements to the auxiliary tapes. Upon reading a blank on all tapes, these computations move to state \texttt{G}, where at some point no transition rules apply and the computations halt. Thus, for these computations there are no new states that need to be highlighted in crimson. In our experience, it is very difficult for many students to understand this explanation without the aid of an \fsm{} \mttm{} computation graph.

Consider applying \texttt{EQABC-ND} to  \quot{}\texttt{(\_ b c a)} using a cut off threshold of 8. The computation graph for this application is displayed in \Cref{eqabc-nd-cutoff}. We can immediately observe that the only state highlighted in crimson, \texttt{G}, is not the accepting state and, given its gold highlighting, that there are computations cut off at state \texttt{G}. Thus, no accepting computations were found, but we also cannot conclude anything about the word being in \texttt{EQABC-ND}'s language. To understand the possible computations, once again, we use the \mttm{} computation graph to reason about the phases. As in the example above, there is a computation that moves from \texttt{S} to \texttt{G}, where no transition rules apply and the computation halts. Thus, explaining why \texttt{G} is highlighted in crimson. As above, there are also computations that reach phase 2 and copy the input word's elements to the auxiliary tapes. We know there is more than one such computation, because multiple rules for each alphabet element are used to copy a single instance of each. When these computations transition to state \texttt{G} to start phase 3, they are cut off given that the threshold has been reached. Thus, \texttt{G} is rendered using gold. To better understand why all these computations are cut off, constructing the trace of a single computation based on the \mttm{} computation graph usually suffices. We reconstruct a computation that few, if any, students realize is possible:
\begin{alltt}
     (S (1 0 0 0) ((@ _ a b c)(_)(_)(_))) 
       \step (C (2 1 1 1) ((@ _ a b c)   (_ _)       (_ _)  (_ _)))
       \step (D (2 1 1 1) ((@ _ a b c)   (_ a)       (_ _)  (_ _)))
       \step (C (3 2 1 1) ((@ _ a b c)   (_ a _)     (_ _)  (_ _)))
       \step (D (3 2 1 1) ((@ _ a b c)   (_ a b)     (_ _)  (_ _)))
       \step (C (4 3 1 1) ((@ _ a b c)   (_ a b _)   (_ _)  (_ _)))
       \step (D (4 3 1 1) ((@ _ a b c)   (_ a b c)   (_ _)  (_ _)))
       \step (C (5 4 1 1) ((@ _ a b c _) (_ a b c _) (_ _)  (_ _)))
       \step (G (4 3 0 0) ((@ _ a b c _) (_ a b c _) (_ _)  (_ _)))                             
\end{alltt}
In this computation, all word elements on \texttt{t0} are copied to \texttt{t1}. In \texttt{G}, the threshold is reached and the computation is cut off. Many students do not realize this is a possible computation, because at first they do not realize that the machine does not have to use every tape. Once this example is understood, it is not difficult for students to understand that the machine is capable of copying word elements to the auxiliary tapes in any manner. We note that increasing the threshold to 10 allows for computations to reach \texttt{Y} and accept before reaching the cut off threshold.

Finally, consider applying \texttt{EQABC-ND} to  \quot{}\texttt{(\_ b c a a c b)} and its computation graph displayed in \Cref{eqabc-nd-accept}. It is straightforward to see why the machine accepts. Observe that \texttt{a} is always copied to \texttt{t2}, given that there is a single transition rule on an \texttt{a} from \texttt{D} to \texttt{C}. For the two instances of \texttt{b}s and of \texttt{c}s, observe that for each one is copied to tape 1 and one to tape 3 (again, based on the transitions rules from \texttt{D} to \texttt{C}). We are free to choose which instance is copied to which tape to reconstruct the following accepting computation:
\begin{alltt}
     (S (1 0 0 0) ((@ _ b c a a c b)(_)(_)(_))) 
       \step (C (2 1 1 1) ((@ _ b c a a c b)    (_ _)      (_ _)      (_ _)))
       \step (D (2 1 1 1) ((@ _ b c a a c b)    (_ b)      (_ _)      (_ _)))
       \step (C (3 2 1 1) ((@ _ b c a a c b)    (_ b _)    (_ _)      (_ _)))
       \step (D (3 2 1 1) ((@ _ b c a a c b)    (_ b c)    (_ _)      (_ _)))
       \step (C (4 3 1 1) ((@ _ b c a a c b)    (_ b c _)  (_ _)      (_ _)))
       \step (D (4 3 1 1) ((@ _ b c a a c b)    (_ b c _)  (_ a)      (_ _)))
       \step (C (5 3 2 1) ((@ _ b c a a c b)    (_ b c _)  (_ a _)    (_ _)))
       \step (D (5 3 2 1) ((@ _ b c a a c b)    (_ b c _)  (_ a a)    (_ _)))
       \step (C (6 3 3 1) ((@ _ b c a a c b)    (_ b c _)  (_ a a _)  (_ _)))
       \step (D (6 3 3 1) ((@ _ b c a a c b)    (_ b c _)  (_ a a _)  (_ c)))
       \step (C (7 3 3 2) ((@ _ b c a a c b)    (_ b c _)  (_ a a _)  (_ c _)))
       \step (D (7 3 3 2) ((@ _ b c a a c b)    (_ b c _)  (_ a a _)  (_ c b)))
       \step (C (8 3 3 3) ((@ _ b c a a c b _)  (_ b c _)  (_ a a _)  (_ c b _)))
       \step (G (8 2 2 2) ((@ _ b c a a c b _)  (_ b c _)  (_ a a _)  (_ c b _)))
       \step (G (8 1 1 1) ((@ _ b c a a c b _)  (_ b c _)  (_ a a _)  (_ c b _)))
       \step (Y (8 1 1 1) ((@ _ b c a a c b _)  (_ b c _)  (_ a a _)  (_ c b _)))                             
\end{alltt}
In this reconstruction, the first \texttt{b} is copied to tape 1, the second \texttt{b} is copied to tape 3 and the same is done for the \texttt{c}s.

This concludes our presentation of computation graphs for nondeterministic \mttm{}s. In our experience, we have found that explaining the observed behavior of nondeterministic \mttm{}s is made easier with their use. We note, however, that an \mttm{} computation graph does not guarantee that understanding is forthcoming. To be effective, this tool does require the \fsm{} programmer to have a good grasp of the machine's design, including the role of each state and the phases the machine operates in. Otherwise, the transitions carry little meaning making the \mttm{} computation graph difficult to understand.

\section{Student Feedback and Limitations}
\label{data}

\subsection{Empirical Data}

This section presents empirical data collected from undergraduate students in a \flatt{} course at Seton Hall University that uses the described visualization tools for \mttm{} education. The course enrolled 11 students of which 10 volunteered to participate in the study. All participants are third or fourth year Computer Science majors (3 females and 7 males), taking their first \flatt{} course, none had any prior experience with \mttm{}s, and none received any benefit or compensation for participating in the study.

The course's \mttm{} module consists of 2 75-minute lectures presenting the design, implementation, and verification of several \mttm{}s. Students had a week to complete assignments that included, for example, the design of an \mttm{} to decide  \texttt{L=a$^{\texttt{n}}$b$^{\texttt{n}}$c$^{\texttt{n}}$}. Throughout the lectures the visualization tools described are used and students are strongly encouraged to use the tools while solving assigned problems.

We empirically evaluate student experiences using an anonymous survey consisting of 9 questions answered using a Likert scale \cite{Likert}: 4 focusing on \mttm{} transition diagrams and 5 focusing on \mttm{} computation graphs. Each question presents a statement and respondents are asked to indicate their level of agreement using the scale: 1 (Strongly Disagree) to 5 (Strongly Agree), including, 3, a neutral category. 

\begin{figure}[t!]
\begin{tikzpicture}[scale=0.80]
\begin{axis}[
    bar width=.5cm,
    width=1.2\textwidth,
    height=.3\textheight,
    ybar,
    enlargelimits=0.15,
    legend style={at={(0.5,1.3)},
      anchor=north,legend columns=-1},
    ylabel={Proportion of Respondents},
    symbolic x coords={1,2,3,4,5},
    xlabel={1 (Strongly Disagree)\ldots{}5 (Strongly Agree)},
    xtick=data,
    nodes near coords,
    nodes near coords align={vertical},
    ]
\addplot[color=black,fill=blue!50] coordinates {(1,0) (2,0) (3,0) (4,.10) (5,.90)};
\addplot[color=black,fill=blue!] coordinates {(1,0) (2,0) (3,0) (4,.10) (5,.90)};
\addplot[color=black,fill=heliotrope] coordinates {(1,0) (2,0) (3,.10) (4,.10) (5,.80)};
\addplot[color=black,fill=pink] coordinates {(1,0) (2,0) (3,0) (4,.10) (5,.90)};
\legend{Q1, Q2, Q3, Q4}
\end{axis}
\end{tikzpicture}
\caption{Data distribution for questions on \mttm{} transition diagrams.} \label{td-data}
\end{figure}
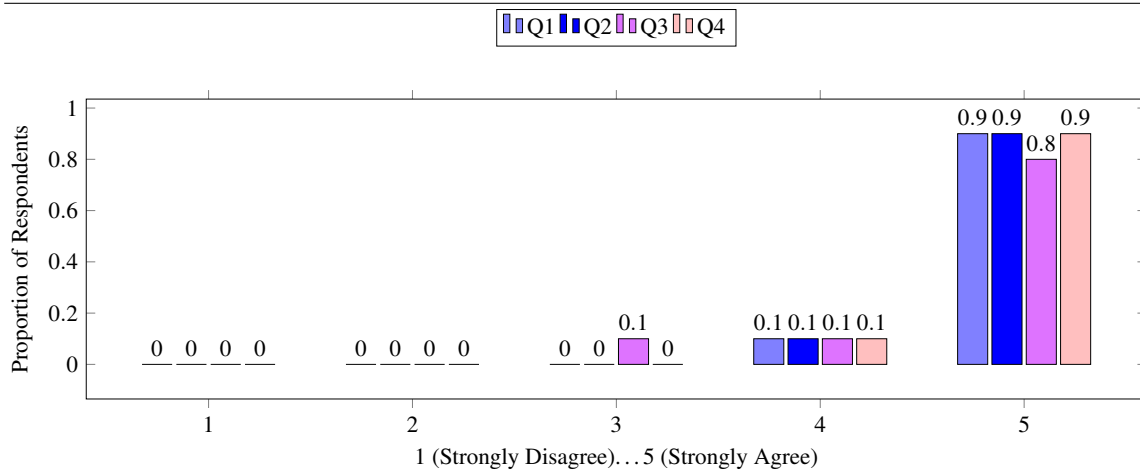

Regarding \mttm{} transition diagrams, the survey presents the following statements: 
\begin{alltt}
  \textcolor{blue!50}{Q1}: Transition diagrams are useful to understand \mttm{}s.
  \textcolor{blue}{Q2}: Transition diagrams clearly display the transition relation.
  \textcolor{heliotrope}{Q3}: Transition diagrams are visually appealing.
  \textcolor{pink}{Q4}: The coloring scheme and use of shapes is useful and clear.
\end{alltt}
\Cref{td-data} presents the data distribution. For \textcolor{blue!50}{Q1}, we observe that all respondents tend to agree (i.e., responses 4 and 5) with the statement. This suggests that the students feel strongly about \mttm{} transition diagrams being useful to understand these machines. For \textcolor{blue}{Q2}, all participants tend to agree with the statement as evidenced by all responses given are 4 or 5. Given that we were worried about the edge labels being too cumbersome, this is a very encouraging result that suggests that students find the rendering of the transition rules clear in the produced graphics. For \textcolor{heliotrope}{Q3}, we observe that an overwhelming majority of respondents, 90\%, feel that the transition diagrams for \mttm{}s are visually appealing (i.e., responses 4 and 5) and 10\% feel neutral about this characteristic (response 3). This result is also encouraging because in our experience students are more likely to use and learn from graphics that are visually appealing to them. Finally, for \textcolor{pink}{Q4}, we observe that all students tend to agree that the coloring scheme and use of shapes is useful and clear (i.e., responses 4 and 5). 

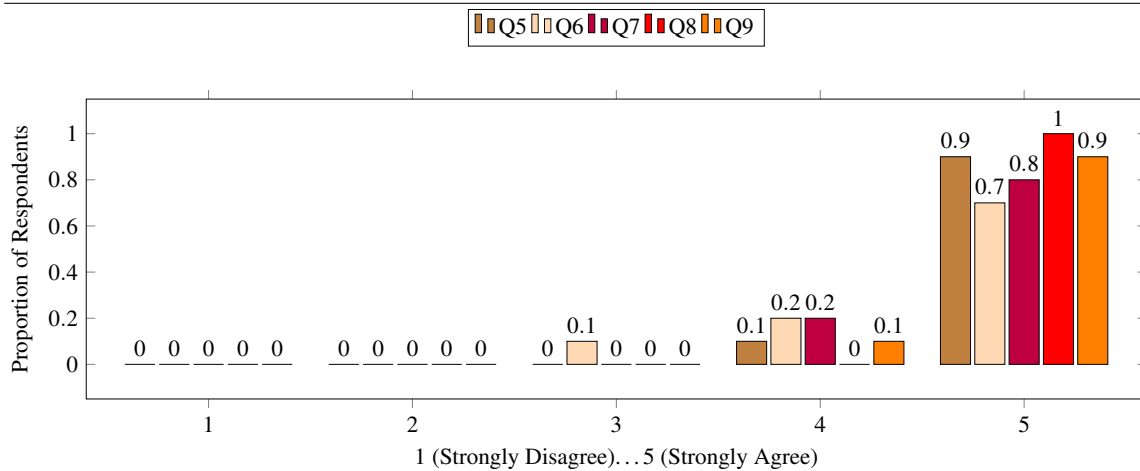
\begin{figure}[t!]
\begin{tikzpicture}[scale=0.80]
\begin{axis}[
    bar width=.5cm,
    width=1.2\textwidth,
    height=.3\textheight,
    ybar,
    enlargelimits=0.15,
    legend style={at={(0.5,1.3)},
      anchor=north,legend columns=-1},
    ylabel={Proportion of Respondents},
    symbolic x coords={1,2,3,4,5},
    xlabel={1 (Strongly Disagree)\ldots{}5 (Strongly Agree)},
    xtick=data,
    nodes near coords,
    nodes near coords align={vertical},
    ]
\addplot[color=black,fill=brown] coordinates {(1,0) (2,0) (3,0) (4,.10) (5,.90)};
\addplot[color=black,fill=orange!30] coordinates {(1,0) (2,0) (3,.10) (4,.20) (5,.70)};
\addplot[color=black,fill=purple] coordinates {(1,0) (2,0) (3,0) (4,.20) (5,.80)};
\addplot[color=black,fill=red] coordinates {(1,0) (2,0) (3,0) (4,0) (5,1.00)};
\addplot[color=black,fill=orange] coordinates {(1,0) (2,0) (3,0) (4,.10) (5,.90)};
\legend{Q5, Q6, Q7, Q8, Q9}
\end{axis}
\end{tikzpicture}
\caption{Data distribution for questions on \mttm{} computation graphs.} \label{cg-data}
\end{figure}

Regarding \mttm{} computation graphs, the survey presents the following statements: 
\begin{alltt}
  \textcolor{brown}{Q5}: Computation graphs are useful to understand \mttm{}s.
  \textcolor{orange!50}{Q6}: Computation graphs help explain multiple \mttm{} computations on a word.
  \textcolor{purple}{Q7}: Computation graphs clearly summarize all possible \mttm{} computations.
  \textcolor{red}{Q8}: The role of cutoff thresholds is clear.
  \textcolor{orange}{Q9}: The coloring scheme is clear.
\end{alltt}
The distribution of responses is displayed in \Cref{cg-data}. For \textcolor{brown}{Q5}, all respondents tend to agree with the statement (i.e., responses 4 and 5). This suggests that \mttm{} computation graphs have met our expectations to help students better understand how \mttm{}s operate and why they accept/reject. For \textcolor{orange!50}{Q6}, we observe that 90\% of the respondents tend to agree (i.e., responses 4 and 5) that computation graphs are useful to understand why \mttm{}s have multiple computations on a word and 10\% tend to feel neutral (i.e., response 3). Given that the respondents are introduced to nondeterminism for the first time in this course, it is expected for some students to feel less strongly. Nonetheless, the distribution of responses suggests that \mttm{} computation graphs are effective in helping students understand nondeterminism. For \textcolor{purple}{Q7}, we observe that  all respondents feel strongly that computation graphs clearly summarize all possible computations (i.e., responses 4 and 5). This further validates that we have achieved our goal to help students better understand nondeterminism. For \textcolor{red}{Q8}, all respondents tend to strongly agree (i.e., response 5) that the role of cutoff thresholds for \mttm{} computation graphs is clear. This suggests that, in general, respondents understand that \mttm{}s may not halt and, therefore, computations may need to be cut off. Finally, for \textcolor{orange}{Q9}, we observe that all respondents tend to agree that the coloring scheme is clear (responses 4 and 5). This suggest that the chosen color scheme is helpful when interpreting \mttm{} computation graphs.

\subsection{Limitations}

We acknowledge that there are several threats to validity and we briefly highlight a few. The first is that our sample size is small. This means that it is unclear if similar results would be obtained from a larger sample. A larger sample is likely to include students with more diverse programming and mathematical abilities. The second is that our sample comes from a single university located in northeast USA in which the majority of Computer Science students are male, European Americans (i.e., the largest panethnic group in the U.S.), and of medium to high social-economic status--a factor known to influence achievement in Computer Science \cite{Haitte,Parker,Sirin,White2}. This may limit the applicability of our observations to populations in different geographical areas with a different diversity mix. Finally, the data was collected from students taught by a single instructor. Therefore, there may be confounding factors, like instructor approval, that may have had an impact on student responses.

To address these potential pitfalls, future work includes more in-depth studies over several iterations of the course both at Seton Hall and at other institutions that adopt our approach. In this manner, we expect to collect data from a more diverse student sample from different regions of the world being taught by different instructors.

\section{Related Work}
\label{rw}

Most \texttt{FLAT} textbooks that introduce \mttm{}s (e.g., \cite{Hopcroft,Lewis,Linz,Rich,Sipser}) rely only on static graphics to help readers understand the operational semantics of \mttm{}s. These graphics illustrate a machine's transition diagram or illustrate head movements and tape mutations. Little emphasis, if any, is placed on design, on helping readers understand nondeterminism, and on why an \mttm{} accepts or rejects. None mention nor use computation graphs. This last limitation is understandable, because explaining how to generate a computation graph without automation is likely to be time-consuming and difficult to cover in most \flatt{} courses. As done in most textbooks that discuss \mttm{}s, the approach presented in this article also relies on graphics. In contrast, however, the work presented in this article aims to directly help students design and understand why \mttm{}s accept or reject. It does so by automatically generating transition diagrams and computation graphs for any \mttm{} a student develops. Furthermore, \fsm{} provides a dynamic visualization tool to simulate machine execution. Such an approach goes beyond what is possible in a textbook by providing students with feedback on their own designs.

\begin{figure}[t!]
\centering
\includegraphics[scale=0.45]{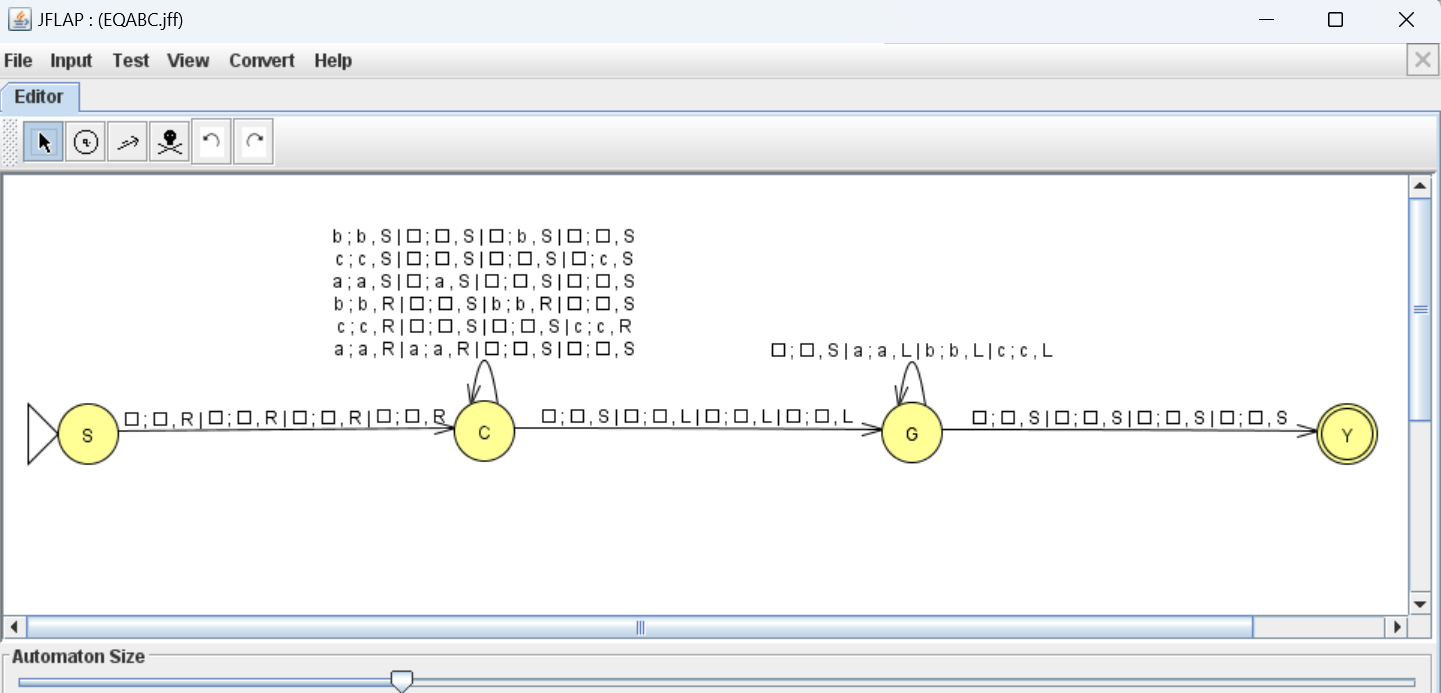}
\caption{The \mttm{} from \Cref{eqabc-nd-td} in \texttt{JFLAP}.} \label{eqabc-nd-td-jflap}
\end{figure}

The best-known graphical tools to support \flatt{} instruction are \texttt{JFLAP} \cite{Rodger,RodgerII,RodgerIII} and \texttt{Automata Tutor} \cite{DAntoni2,DAntoni}. \texttt{JFLAP} limits the number of \mttm{} tapes to a maximum of 5 and requires the user to choose a visually appealing layout \cite{jflap-tutorial}. In our experience, students are distracted from \mttm{} design by tinkering to find a graphical layout that satisfies them. In further contrast with \fsm{}, transitions always specify two actions for each tape: a tape mutation and a head movement. The transitions are rendered as edges whose labels are organized by tapes: each transition label contains the actions for each tape separately listed. For instance, \Cref{eqabc-nd-td-jflap} displays an implementation of \texttt{EQABC-ND} from \Cref{eqabc-nd-td}. Observe that state \texttt{D} is not needed, because tape mutations and head movements are performed by a single transition. This reduction in the number of states comes at a cost. First, the \mttm{} designer must reason about two actions in one transition, which makes developing correctness arguments harder. Second, the graphic is not as visually appealing nor as clear as those produced by \fsm{}. For instance, the labels on the edges tend to be much longer and some effort is required to ascertain the actions taken on each tape for a given transition. This contrasts sharply with \fsm{} \mttm{} graphs, which render transitions only using two parts: the read elements on each tape and the action taken on each tape. Finally, we note that \texttt{JFLAP} does not support computation graphs.

In \texttt{Automata Tutor}, all Turing machines may have multiple tapes. A standard Turing machine, for instance, is defined as having a single tape. In contrast to the approach taken using \fsm{}, \texttt{Automata Tutor} does not focus on the systematic design of machines. Instead, it focuses on providing counterexamples and conceptual hints to help students successfully implement machines. In further contrast, suggested exercises focus on converting a \texttt{while}-loop into a Turing machine. Finally, neither machine verification nor computation graphs are supported by \texttt{Automata Tutor}.

\section{Conclusions and Future Work}
\label{concls}

This article presents the visual support provided by \fsm{}, a \texttt{DSL} embedded in \racket{} for the Automata Theory classroom, to help students design and understand multitape Turing machines. Three tools are presented: a dynamic visualization tool to simulate machine execution, a static visualization tool to generate \mttm{} transition diagrams and phase-based transition diagrams, and a static visualization tool to generate \fsm{} \mttm{} computation graphs. The dynamic visualization tool allows users to apply any \mttm{}, deterministic or nondeterministic, they have implemented to a word in the machine's language and observe, both forwards and backwards, the steps taken in the computation. The generation of transition diagrams and phase-based transition diagrams is done using \gviz{} to make the graphics visually appealing. The static visualization tool to generate \fsm{} \mttm{} computation graphs also uses \gviz{} to make the graphics visually appealing. In addition, it uses color coding and node shapes to highlight where computations end, where computations are cut off, and the different types of states (i.e., ordinary, starting, final, and accepting). Empirical data collected from students in a \flatt{} course at Seton Hall University using these tools, suggests that the visualizations are well-received, clear, and useful to understand \mttm{}s.

Future work includes a larger more in-depth empirical study, across several offerings of the course, to better understand how students in a classroom setting feel about the developed visualization tools and to determine how they may be improved. In addition, inspired by the work presented in this article, we are exploring how to graphically summarize all the possible derivations performed by a context-free grammar that has not been transformed to Chomsky normal form \cite{Chomsky} nor to Greibach normal form \cite{Greibach}.

\balance
\bibliographystyle{eptcs}
\bibliography{mttm-trans-diag-cmp-graph}

\newpage

\appendix

\section{Implementation for \texttt{EQABC-ND}}
\label{A1}

\begin{figure}[h!]
\begin{lstlisting}[language=racket,escapechar=\%]
(define EQABC-ND
  (make-mttm
    %\quot{}%(S Y C D G) %\quot{}%(a b c) %\quot{}%S %\quot{}%(Y)
    (list
      (list %\quot{}%(S (_ _ _ _))  %\quot{}%(C (R R R R)))
      (list '(S (_ _ _ _))  '(G (R R R R)))
      ;; copy an a to any tape
      (list %\quot{}%(C (a _ _ _))  %\quot{}%(D (a a _ _)))
      (list %\quot{}%(D (a a _ _))  %\quot{}%(C (R R _ _)))
      (list %\quot{}%(C (a _ _ _))  %\quot{}%(D (a _ a _)))
      (list %\quot{}%(D (a _ a _))  %\quot{}%(C (R _ R _)))
      (list %\quot{}%(C (a _ _ _))  %\quot{}%(D (a _ _ a)))
      (list %\quot{}%(D (a _ _ a))  %\quot{}%(C (R _ _ R)))
      ;; copy a b to any tape
      (list %\quot{}%(C (b _ _ _))  %\quot{}%(D (b b _ _)))
      (list %\quot{}%(D (b b _ _))  %\quot{}%(C (R R _ _)))
      (list %\quot{}%(C (b _ _ _))  %\quot{}%(D (b _ b _)))
      (list %\quot{}%(D (b _ b _))  %\quot{}%(C (R _ R _)))
      (list %\quot{}%(C (b _ _ _))  %\quot{}%(D (b _ _ b)))
      (list %\quot{}%(D (b _ _ b))  %\quot{}%(C (R _ _ R)))
      ;; copy a c to any tape
      (list %\quot{}%(C (c _ _ _))  %\quot{}%(D (c c _ _)))
      (list %\quot{}%(D (c c _ _))  %\quot{}%(C (R R _ _)))
      (list %\quot{}%(C (c _ _ _))  %\quot{}%(D (c _ c _)))
      (list %\quot{}%(D (c _ c _))  %\quot{}%(C (R _ R _)))
      (list %\quot{}%(C (c _ _ _))  %\quot{}%(D (c _ _ c)))
      (list %\quot{}%(D (c _ _ c))  %\quot{}%(C (R _ _ R)))
      ;; match as, bs, and cs
      (list %\quot{}%(C (_ _ _ _))  %\quot{}%(G (_ L L L)))
      (list %\quot{}%(G (_ a b c))  %\quot{}%(G (_ L L L)))
      (list %\quot{}%(G (_ a c b))  %\quot{}%(G (_ L L L)))
      (list %\quot{}%(G (_ b a c))  %\quot{}%(G (_ L L L)))
      (list %\quot{}%(G (_ b c a))  %\quot{}%(G (_ L L L)))
      (list %\quot{}%(G (_ c a b))  %\quot{}%(G (_ L L L)))
      (list %\quot{}%(G (_ c b a))  %\quot{}%(G (_ L L L)))
      (list %\quot{}%(G (_ _ _ _))  %\quot{}%(Y (_ _ _ _))))
    4
    %\quot{}%Y))
\end{lstlisting}
\caption{A nondeterministic \mttm{} implementation for \texttt{L}=\{\texttt{w} $|$ \texttt{w} has equal number of \texttt{a}s, \texttt{b}s, and \texttt{c}s\}.} \label{rem-defs}
\end{figure}

\end{document}